\documentclass[10pt]{article}
\usepackage{amssymb}
\usepackage{graphicx}
\def\N{\mathbb{N}}

\def\kleq{\simeq}

\def\lsem{[\![}
\def\rsem{]\!]}
\newcommand\sem[1]{\lsem #1 \rsem}

\def\iso{\cong}

\def\AA{{\mathbf{A}}}

\def\TT{{\mathbf{T}}}

\def\rlzn{\mbox{$\,-\!\!\!-\!\!\rhd\,$}}
\def\arrow{\rightarrow}
\def\parrow{\rightharpoonup}

\def\darrow{\Rightarrow}

\def\reducesto{\rightsquigarrow}

\def\Set{{\mathcal{S}\!\!et}}

\def\id{{\mathit{id}}}

\newcommand\num[1]{\widehat{#1}}  

\def\dom{{\mathrm{dom}}}

\def\Fst{{\mathit{fst}}}
\def\Snd{{\mathit{snd}}}

\def\str{{\mathit{str}}}

\def\Tsf{{\mathsf{T}}}

\def\Gcal{{\mathcal{G}}}

\def\nat{{\mathtt{N}}}  

\def\iif{{\mathit{if}}}
\def\tthen{{\mathit{then}}}
\def\eelse{{\mathit{else}}}

\def\ccase{{\mathtt{case}}}
\def\oof{{\mathtt{of}}}

\newcommand\caseof[2]{\ccase\;#1\;\oof\;(#2)}
\newcommand\ang[1]{\langle #1 \rangle}

\newcommand\myvec[1]{\vec{#1}}

\def\suc{{\mathit{suc}}}
\def\pre{{\mathit{pre}}}

\def\rec{{\mathit{rec}}}
\def\Rec{{\mathit{Rec}}}
\def\ifzero{{\mathit{ifzero}}}

\def\mmin{{\mathit{min}}}
\def\byval{{\mathit{byval}}}

\def\Set{{\mathsf{S}}}

\def\Ct{{\mathsf{Ct}}}

\def\HEO{{\mathsf{HEO}}}

\def\PC{{\mathsf{PC}}}

\def\SP{{\mathsf{SP}}}

\def\SF{{\mathsf{SF}}}

\def\Kl{{\mathsf{Kl}}}
\def\Klex{{\mathsf{Klex}}}

\def\eff{{\mbox{\scriptsize \rm eff}}}
\def\obs{{\mbox{\scriptsize \rm obs}}}

\def\lwf{{\mbox{\scriptsize \rm lwf}}}
\def\lbd{{\mbox{\scriptsize \rm lbd}}}
\def\basic{{\mbox{\scriptsize \rm basic}}}

\def\prim{{\mbox{\scriptsize \rm prim}}}
\def\smallmin{{\mbox{\scriptsize \rm min}}}

\def\PCF{{\mathrm{PCF}}}
\def\sysT{{\mathrm{T}}}

\def\LLL{{\mathcal{L}}}

\def\BR{{\mathsf{BR}}}

\newcommand\details[1]{}

\newcommand\pending[1]{}

\newtheorem{definition}{Definition}
\newtheorem{proposition}[definition]{Proposition}
\newtheorem{lemma}[definition]{Lemma}
\newtheorem{theorem}[definition]{Theorem}

\def\proof{{\sc Proof:}~}
\def\proofsk{{\sc Proof sketch:}~}
\def\QED{$\Box$}

\def\while{\mathit{while}}
\def\fst{\Fst}
\def\snd{\Snd}
\def\enc{\mathit{enc}}
\def\dec{\mathit{dec}}

\def\sysW{\mathrm{W}}
\def\sysB{\mathrm{B}}
\def\obseq{\simeq_\obs}
\def\TT{\mathcal{T}}
\def\len{\mathrm{len}}
\def\append{\mathrm{add}}
\def\basic{\mathrm{basic}}
\def\nf{\mathrm{nf}}

\begin{document}

\title{Bar recursion is not computable via iteration}
\author{John Longley}
\maketitle

\begin{abstract}
We show that the \emph{bar recursion} operators of Spector and Kohlenbach,
considered as third-order functionals acting on total arguments, are not computable 
in G\"odel's System T plus minimization, which we show to be equivalent to 
a programming language with a higher-order iteration construct.
The main result is formulated so as to imply the non-definability of bar recursion in $\sysT+\min$ 
within a variety of partial and total models, for instance the Kleene-Kreisel continuous functionals.
The paper thus supplies proofs of some results stated in the book by Longley and Normann.

The proof of the main theorem makes serious use of the theory of \emph{nested sequential procedures}
(also known as PCF B\"ohm trees), and proceeds by showing that bar recursion cannot be
represented by any sequential procedure within which the tree of nested function applications is well-founded.
\end{abstract}

\section{Introduction}  \label{sec-intro}

In the study of computability theory in a higher-order setting, 
where `computable operations' may themselves be passed as arguments to other computable operations,
considerable interest attaches to questions of the relative power of different programming languages
or other formalisms for computation \cite{HOC}.
In this paper, we shall compare the expressive power of 
a higher-order language supporting general \emph{iteration} (in the sense of \textit{while} loops) 
with one supporting general \emph{recursion} (as in recursive function definitions).

On the one hand, it will be easy to see that our iteration constructs are definable via recursion,
so that the second language subsumes the first.
On the other hand, there is an example due to Berger \cite{Berger-min-rec} of a second-order functional $H$,
informally of type $(\N_\bot \times \N_\bot \arrow \N_\bot) \arrow \N_\bot$,
which is definable via recursion but not via iteration (see Section~\ref{sec-history} below).
So in this sense at least, we may already say that iteration is weaker than recursion.

However, it is crucial to Berger's example that we are considering the behaviour of $H$ on
arbitrary (hereditarily) \emph{partial} arguments rather than just on total ones:
indeed, Berger also showed that if we merely ask which functionals of types
$(\N^r \arrow \N) \arrow \N_\bot$ are representable,
then iteration (even in a weak form) turns out to be just as powerful as recursion.
One may therefore wonder whether, more generally, iteration and recursion offer equally
powerful means for defining operations on `hereditarily total' arguments.
The question is a natural one to ask in a computer science context, since it has sometimes been
suggested that it is only the behaviour of a program on total arguments that is likely to matter
for practical purposes (see Plotkin \cite{Plotkin-totality}).

The main contribution of this paper is to answer this question in the negative: 
at third order, there are `hereditarily total' functionals definable by very simple kinds recursion, 
but not by even the most general kind of iteration that we can naturally formulate.
Indeed, one example of such a functional is the well-known \emph{bar recursion} operator, 
first introduced by Spector in the context of interpretations of classical analysis \cite{Spector-bar}.
Since bar recursion and its close relatives themselves offer a number of intriguing programming possibilities
that are active topics of current research
(e.g.\ within game theory \cite{Nash-equilibria,HO-decision-theory} 
and proof mining \cite{Kohlenbach-book,Ramsey-BR,Podelski-Rybalchenko-BR}),
we consider this to be an especially significant example of 
the expressivity difference between iteration and recursion.

More specifically,
we will show that neither Spector's original bar recursion functional nor the variant due to Kohlenbach 
\cite{Kohlenbach-bar-rec} is computable in a language with `higher-order iteration', 
even if we restrict attention to `hereditarily total' arguments.
As we shall see, there is more than one way to such a statement precise,
but we shall formulate our theorem in a robust form which (we shall argue) 
establishes the above claim in all reasonable senses of interest.%
\footnote{The main results of this paper were stated in \cite{HOC} 
as Theorem 6.3.28 and Corollary 6.3.33, with a reference to a University of Edinburgh technical report
\cite{Bar-rec-not-Tmin} for the proof. The present paper is a considerably reworked and expanded version 
of this report, incorporating some minor corrections, and more fully developing the connection with 
familiar iteration constructs (hence the change in the title).}

As our framework for computation with recursion, we shall work with Plotkin's well-known language 
PCF for partial computable functionals \cite{LCF-considered}, in which recursion is embodied by
a \emph{fixed point} operator $Y_\sigma: (\sigma \arrow \sigma) \arrow \sigma$ for each type $\sigma$.
For iteration, we shall introduce a bespoke language $\sysW$ with a higher-order \textit{while} construct,
and show that it is equivalent in power to G\"odel's System~T extended with the familiar 
minimization (i.e.\ unbounded search) operator $\min$.
In fact, both $\PCF$ and $\sysT+\min$ have precursors and analogues in the earlier
literature on higher-order computability, and the study of the relationship between 
(broadly) `recursive' and `iterative' styles of computation turns out to have quite deep historical roots.
We now survey some of this history in order to provide some further context for our present work.

\subsection{Historical context}  \label{sec-history}

In a landmark paper of 1959, Kleene \cite{RFQFT-I} provided the first full-blown generalization of a
concept of `effective computability' to all finite type levels, working within the full set-theoretic type
structure $\Set$ of hereditarily total functionals over $\N$.
This consisted of an inductive definition of computations via nine schemes S1--S9,
resulting in the identification of a substructure $\Set^\Kl \subset \Set$ consisting of what we now call 
the \emph{Kleene computable} functionals (Kleene himself called them \emph{general recursive}).
Kleene's scheme S9, in particular, postulates in effect the existence of a `universal' computable functional,
and this in turn gives rise to a very general form of recursive function definition 
(see e.g.\ \cite[Section~5.1.2]{HOC}).
Indeed, although Kleene's S1--S9 definition looks superficially very different from Plotkin's PCF,
it turns out that in a certain sense, the two formalisms express exactly the same class of algorithms for 
higher-order computation (see \cite[Sections~6.2 and 7.1]{HOC}).%
\footnote{Strictly speaking, to obtain this equivalence at the algorithmic level, we need a mild extension
of PCF with an operator $\byval$ as described in Subsection~\ref{sec-NSPs} below.}

In the same paper, Kleene also considered another notion of computability in which S9 was replaced by a 
weaker scheme S10 for minimization (= unbounded search),
giving rise to a substructure $\Set^\smallmin \subset \Set^\Kl$ of \emph{$\mu$-computable} functionals
(Kleene's terminology was \emph{$\mu$-recursive}). 
Whereas we can regard S9 as giving us `general recursion', it is natural to think of S10 as giving us
a particularly simple kind of `iteration': indeed, from a modern perspective, we may say that S1--S8 + S10 
corresponds to a certain typed $\lambda$ calculus $\sysW_0^\str$ with \emph{strict ground-type iteration}, 
or equivalently to a language $\sysT_0^\str+\min$
with strict ground-type primitive recursion and minimization. 

With the spectacles of hindsight, then, we can see that in \cite{RFQFT-I} the stage was already set for a comparison between `iterative' and `recursive' flavours of higher-order computation.
Indeed, in \cite[Section~8]{RFQFT-I}, Kleene showed (in effect) that the System~$\sysT$ recursor 
$\rec_{\nat\arrow\nat}$ (a third-order functional in $\Set$) was Kleene computable but not $\mu$-computable.
However, Kleene's proof relied crucially on the possibility of applying $\rec_{\nat\arrow\nat}$ to
`discontinuous' arguments (in particular the second-order functional $\exists^2$ embodying 
quantification over $\N$); it thus left open the question of whether every $\Psi \in \Set^\Kl$ 
could be mimicked by some $\Psi' \in \Set^\smallmin$ if one restricted attention to `computable' arguments.

Over the next two decades, much of the focus of research shifted from the full set-theoretic model $\Set$
to the Kleene-Kreisel type structure $\Ct$ of \emph{total continuous functionals}, a realm of functionals
of a more `constructive' character than $\Set$ which was found to be better suited to many
metamathematical applications (see e.g.\ \cite{Kreisel-interp-analysis}).
Once again, the notions of $\mu$-computability and Kleene computability respectively pick out
substructures $\Ct^\smallmin \subseteq \Ct^\Kl$ of $\Ct$,
and Kreisel in \cite[page~133]{Kreisel-set-theoretic} explicitly posed the question of whether these coincide.
(The question is harder to answer here than for $\Set$: Kleene's counterexample $\Psi$ can no longer be 
used, because the necessary discontinuous functionals such as $\exists^2$ are no longer present in $\Ct$.)
This question remained open for some years until being answered by 
Bergstra \cite{Bergstra-thesis}, who used an ingenious construction based on the classical theory 
of c.e.\ degrees to produce an example of a third-order functional in $\Ct^\Kl$ but not in $\Ct^\smallmin$.
On the face of it, Bergstra's example seems ad hoc, but one can extract from his argument the fact that---%
once again---the System~T recursor $\rec_{\nat\arrow\nat}$ 
is Kleene computable but not $\mu$-computable (see \cite[Section~8.5.2]{HOC}).
The fact that Kreisel's question remained open for so long in the face of 
such an `obvious' counterexample suggests that non-computability results of this kind 
were not readily accessible to the proof techniques of the time.

Although Bergstra's argument improves on Kleene's in that it does not rely on the presence of 
discontinuous inputs, it still relies on the existence of \emph{non-computable} second-order
functions within $\Ct$. The argument is therefore not as robust as we might like: for example,
it does not establish the non-$\mu$-computability of $\rec_{\nat\arrow\nat}$ within the type structure
$\HEO$ of \emph{hereditarily effective operations}.
For this, the necessary techniques had to await certain developments in the computer science tradition,
which, in contrast to the work surveyed so far, tended to concentrate on type structures of 
hereditarily \emph{partial} functionals rather than total ones.

As far as we are aware, the first study of the relative power of iteration and recursion in a 
partial setting was that of Berger \cite{Berger-min-rec},
who (in effect) compared the languages $\sysT_0 + \min$ and PCF 
in terms of the elements of Scott's well-known model $\PC$ of \emph{partial continuous functionals}
that they define.
Specifically, Berger introduced the partial functional $H \in \PC((\nat^2 \arrow \nat) \arrow \nat)$
defined informally by
\[ H ~=~ \lambda g^{\nat^2 \arrow \nat}.\;g(0,g(1,g(2,g(\cdots)))) \;, \]
and showed that $H$ is readily definable in PCF (using the recursor $Y_{\nat\arrow\nat}$),
but not at all in $\sysT_0 + \min$.

Although Berger concentrated on $\sysT_0 + \min$ definability in $\PC$, his argument suffices 
to show the non-definability of $H$ in the whole of $\sysT + \min$, and also applies when we replace $\PC$ 
by the type structure $\SF^\eff$ of PCF-computable functionals, yielding a slightly stronger result 
(these points are explained in \cite[Section~6.3]{HOC}).
Thus, Berger's result is apparently the first to show that recursion is stronger than iteration in a sense
that might matter to programmers: the behaviour of the functional $H$ cannot be mimicked using iteration alone, even if we restrict attention to \emph{computable} arguments (which we may here take to mean
`arguments definable in $\sysT_0 + \min$').
Note, however, that Berger's example, unlike those of Kleene and Bergstra,
does emphatically depend on the presence of the element $\bot$ in the models.

Berger's paper provided one of the main inspirations for the study of sublanguages of PCF
in the book of Longley and Normann \cite{HOC}.
There, the focus was on modelling the `algorithms' implicit in PCF programs as
\emph{nested sequential procedures} (NSPs), also known as \emph{$\PCF$ B\"ohm trees}.
This is a model that had roots in early work of Sazonov \cite{Sazonov-early}, but which came into focus
in the course of work on game semantics for PCF \cite{AJM-games,HO-games}.
In summary, an NSP is a potentially infinite `decision tree' recording the various function calls 
(including nested calls) that a higher-order program might make, 
along with the dependency of its behaviour on the results of such calls
(further detail will be given in Section~\ref{sec-NSPs}).
One of the main ideas explored in \cite[Chapter~6]{HOC} was that certain sublanguages of PCF can
be correlated with certain classes of NSPs: for instance, any NSP $p$ definable in $\sysT_0^\str + \mmin$
is \emph{left-bounded} (that is, there is a finite global bound $d$ on the nesting depth of function applications
within $p$), while any NSP definable in $\sysT + \mmin$ is \emph{left-well-founded}
(that is, the tree of nested applications within $p$ is well-founded).

It turns out that such observations, together with some concrete combinatorial analysis of NSP computations,
can lead to interesting new non-definability results.
For instance, one of the main new results of \cite{HOC} (Theorem~6.3.27) is that no left-bounded procedure
can have the same behaviour as $\rec_{\nat\arrow\nat}$ when restricted to `total' second-order arguments;
it follows that no program of $\sysT_0^\str + \min$ can faithfully represent the total functional 
$\rec_{\nat\arrow\nat}$ on all `total' computable inputs.
As will become clear below, there is some ambiguity here as regards what `total' ought to mean for NSPs;
however, the theorem in \cite{HOC} was formulated in a robust way so as to be applicable to any reasonable
concept of totality. Moreover, it is also straightforward to transfer the result from NSPs to any \emph{total}
model with appropriate structure---in this way, we obtain a robust statement of the non-$\mu$-computability
of $\rec_{\nat\arrow\nat}$ in such models \cite[Corollary~6.3.33]{HOC}, 
not relying on the presence of non-computable or non-total arguments, 
and immediately applicable to a type structure such as $\HEO$.

This seems to offer a satisfactory conclusion to the story as regards the difference between
$\mu$-computability and more general (PCF or Kleene) computability.
However, one is at this point tempted to ask whether the gap between
these two notions might be closed simply by extending the former with all the System $\sysT$ 
recursors $\rec_\sigma$.
Thus, a revised version of Kreisel's question might read: Is the system $\sysT+\mmin$
as powerful as full Kleene computability for the purpose of defining elements of $\Ct$?
(We have already noted that in partial settings such as Scott's $\PC$, a distinction in power between
$\sysT+\mmin$ and full PCF is established by Berger's $H$ functional.)
Indeed, one may even feel that $\sysT+\mmin$ is the more natural level at which to ask such questions,
especially given that $\sysT+\mmin$ corresponds in expressivity to a language $W$ 
that embodies a very general and natural concept of iteration 
(as represented by \textit{while} loops, possibly manipulating higher-order data).

Our main purpose in this paper is to show that, in fact, the famous \emph{bar recursion} operator
furnishes the desired example of a (third-order) total functional that is
Kleene computable (and hence PCF computable) but not $\sysT+\mmin$ definable.
Put briefly, we shall show that bar recursion does for $\sysT+\mmin$ everything that 
$\rec_{\nat\arrow\nat}$ does for $\sysT_0^\str + \mmin$ as described above.
As a further piece of relevant background, a brief glance at the history of bar recursion is therefore in order.

Whereas the System~$\sysT$ recursors $\rec_\sigma$ allow us to construct functions by recursion on
the natural numbers, bar recursion offers a powerful principle for defining functions by recursion on
well-founded trees (the precise definition will be given in Subsection~\ref{sec-bar-recursors}).
Bar recursion was introduced by Spector \cite{Spector-bar} as a major plank of his
remarkable extension of G\"odel's so-called `Dialectica' interpretation of first-order arithmetic
(modulo a double-negation translation)
to the whole of classical analysis (i.e.\ full second-order arithmetic).
Spector's motivations were thus proof-theoretic: for instance, 
System~$\sysT$ extended with bar recursion offered a language of total functionals
powerful enough to define all functions $\N \arrow \N$ provably total in classical analysis.
Spector's interpretation, and variations on it, 
continue to this day to be a fruitful source of results in applied proof theory \cite{Kohlenbach-book}.

Since System~$\sysT$ itself defines only the provably total functions of first-order arithmetic,
it was thus clear at the outset that bar recursion could not be definable within System~$\sysT$.
However, these ideas are of little help when we move to languages such as $\sysT + \mmin$,
which defines all Turing computable functions---%
nor can methods such as diagonalization be used to establish non-definability in the 
`partial' setting of $\sysT+\mmin$. The results of the present paper thus require quite different techniques.

For the purpose of interpreting classical analysis, one requires versions of bar recursion at many 
different type levels; however, for the purpose of this paper, we may restrict attention to the simplest
non-trivial instance of bar recursion (a third-order operation), since this already turns out to be
non-computable in $\sysT+\mmin$.
Another subtlety concerns the way in which we represent the well-founded tree over which
the recursion takes place. Typically the tree is specified via a functional $F: (\nat\arrow\nat)\arrow\nat$
passed as an argument to the bar recursor---however, different ways of representing trees by such functionals
have turned out to have different proof-theoretic applications. In this paper we shall consider two possible
choices: the one used by Spector, and a variant due to Kohlenbach \cite{Kohlenbach-bar-rec}.
As we shall see, the corresponding versions of bar recursion are actually interdefinable relative to
$\sysT+\mmin$, so that the difference is inessential from the point of view of our main result.

Spector's original treatment of bar recursion was syntactic, but it became clear through work of 
Scarpellini \cite{Scarpellini} and Hyland \cite{Hyland-thesis} that bar recursors could be viewed as
(Kleene computable) functionals within $\Ct$.
Thus, bar recursion was very much in the consciousness of workers in $\Ct$ in the early 1970s,
although it was evidently not obvious at the time that it furnished a rather dramatic example of
a Kleene computable but not $\mu$-computable functional.
We will show in this paper how the more recent perspective offered by nested sequential procedures
helps to make such results accessible.

\subsection{Content and structure of the paper}

The main purpose of the paper is to show that the bar recursion functional $\BR$,
even at the simplest type level of interest and in a somewhat specialized form, 
is not definable in System $\sysT+\mmin$.
As we shall see, there are various choices involved in making this statement precise, 
but our formulation will be designed to be robust with respect to such variations.
Our argument will be closely patterned on the proof of the analogous result for
$\sysT_0^\str+\mmin$ and the System~$\sysT$ recursor $\rec_{\nat\arrow\nat}$ 
(see \cite[Theorem~6.3.27]{HOC}).
However, the present proof will also involve some further twists, illustrating some new possibilities for
reasoning with nested sequential procedures.

In Section~\ref{sec-background} we define the languages mentioned in the above discussion---%
$\PCF$, $\sysT+\min$, $\sysW$ and $\sysT_0^\str + \mmin$---and establish some basic 
relationships between them, in particular showing that $\sysT+\min$ and $\sysW$ are
equivalent in expressive power.
We then summarize the necessary theory of nested sequential procedures (NSPs), relying heavily on
\cite{HOC} for proofs, and in particular introducing the crucial substructure of \emph{left-well-founded}
procedures, which suffices for modelling $\sysT+\min$ and $\sysW$.
We also explain the concepts of (Spector and Kohlenbach) bar recursion that we shall work with.

Section~\ref{sec-main-theorem} is devoted to the proof of our main theorem: 
within the NSP model, no bar recursor can be left-well-founded,
hence no program of $\sysT+\min$ or $\sysW$ can implement bar recursion, even in a weak sense.
As mentioned above, the proof will be closely modelled on the corresponding theorem for $\rec_{\nat\arrow\nat}$: indeed, we shall take the opportunity to explain more fully
certain aspects of that proof that were presented rather tersely in \cite{HOC}.
We shall also explain the new ingredients that form part of the present proof.

In Section~\ref{sec-other-models}, we show how our theorem for NSPs transfers readily to other models,
both partial and total, under relatively mild conditions. As an example, we infer that bar recursion is
not $\sysT+\min$ definable within the type structure $\Ct$ of Kleene-Kreisel continuous functionals.

\section{Definitions and prerequisites}  \label{sec-background}

In this section we summarize the necessary technical background and establish a few preliminary results.
We introduce the languages in question in Subsection~\ref{sec-languages},
the nested sequential procedure model in Subsection~\ref{sec-NSPs},
and bar recursion in Subsection~\ref{sec-bar-recursors}.

\subsection{Some languages for recursion and iteration}  \label{sec-languages}

We start by giving operational definitions of the languages we shall study---
principally $\PCF$, $\sysT+\min$ and $\sysW$---and establishing some basic relationships between them.
A relatively easy result here will be that $\sysT+\min$ and $\sysW$ are equally expressive as sublanguages 
of $\PCF$. Of these, $\sysT+\min$ is of course the more widely known and has served as the vehicle
for previous results in the area (e.g.\ in \cite{HOC}); however, $\sysW$ appears to correspond very directly
to a familiar concept of iteration via \textit{while} loops, suggesting that this level of expressivity is 
a natural one to consider from the perspective of programming language theory.

Our version of $\PCF$ will closely follow that of \cite[Chapter~7]{HOC},
except that we shall also include product types, at least initially.
We shall present all our languages as extensions of a common \emph{base language} $\sysB$.

Specifically, our types $\sigma$ are generated by
\[  \sigma,\tau ~::=~ \nat ~\mid~ \sigma \arrow \tau ~\mid~ \sigma \times \tau \;, \]
Terms of $\sysB$ will be those of the simply typed $\lambda$-calculus (with binary products)
constructed from the constants
\[ \begin{array}{rcll}
  \num{n} & : & \nat & \mbox{for each $n \in \N$} \\
  \mathit{suc},\;\mathit{pre} & : & \nat\arrow\nat \\
  \mathit{ifzero} & : & \nat\arrow\nat\arrow\nat\arrow\nat 
\end{array} \]
Throughout the paper, we shall regard the type of a variable $x$ as intrinsic to $x$, and will
often write $x^\sigma$ to indicate that $x$ carries the type $\sigma$.

We endow $\sysB$ with a call-by-name operational semantics
via the following (small-step) basic reduction rules:
\[ \begin{array}{rclcrcl}
  \fst\,\ang{M,N} & \reducesto & M  & &
  \snd\,\ang{M,N} & \reducesto & N  \\
  (\lambda x.M)N & \reducesto & M[x \mapsto N]    & &
  \mathit{suc}\;\num{n} & \reducesto & \num{n+1}  \\
  \mathit{pre}\;\num{n+1} & \reducesto & \num{n} & &
  \mathit{pre}\;\num{0}   & \reducesto & \num{0} \\
  \mathit{ifzero}\;\num{0} & \reducesto & \lambda xy.x  & &
  \mathit{ifzero}\;\num{n+1} & \reducesto & \lambda xy.y 
\end{array} \]
We furthermore allow these reductions to be applied in certain term contexts.
Specifically, the relation $\reducesto$ is inductively generated by the basic rules above together with the clause:
if $M \reducesto M'$ then $E[M] \reducesto E[M']$,
where $E[-]$ is one of the \emph{basic evaluation contexts}
\[ [-]N      ~~~~~~ \suc\,[-]  ~~~~~~ 
   \pre\,[-]  ~~~~~~ \ifzero\,[-]  ~~~~~~
   \fst\,[-]  ~~~~~~ \snd\,[-] \;. \]

We shall consider extensions of $\sysB$ with operations embodying various principles of
general recursion, iteration, primitive recursion and minimization.
To these we also add certain operations that allow us to pass arguments of type $\nat$ `by value',
which will be needed for technical reasons.
The operations we consider are given as constants
\begin{eqnarray*}
Y_\sigma        & : & (\sigma\arrow\sigma) \arrow \sigma \\
\while_\sigma & : & (\sigma\arrow\nat) \arrow \sigma \arrow (\sigma\arrow\sigma) \arrow \sigma \\
\rec_\sigma    & : & \sigma \arrow (\sigma \arrow \nat \arrow \sigma) \arrow \nat \arrow \sigma \\
\mmin            & : & (\nat\arrow\nat) \arrow \nat \arrow \nat \\
\byval^{\vec{\sigma}}_\tau  & : & (\vec{\sigma} \arrow \nat \arrow \tau) \arrow (\vec{\sigma} \arrow \nat \arrow \tau)
\end{eqnarray*}
(where $\vec{\sigma}\arrow\rho$ abbreviates $\sigma_0 \arrow\cdots\arrow \sigma_{r-1} \arrow \rho$
if $\vec{\sigma} = \sigma_0,\ldots,\sigma_{r-1}$),
with associated basic reduction rules
\begin{eqnarray*}
Y_\sigma \; F & \reducesto & F(Y_\sigma F) \\
\while_\sigma\;C\;X\;F & \reducesto & \ifzero\;(CX)\;(\while_\sigma\;C\;(FX)\;F) \; X \\
\rec_\sigma\;X\;F\;\num{0}     & \reducesto & X \\
\rec_\sigma\;X\;F\;\num{n+1} & \reducesto & F\,(\rec_\sigma\,X\,F\,\num{n})\,\num{n} \\
\mmin\;F\;\num{n}                 & \reducesto & 
           \ifzero\;(F\,\num{n})\;\num{n}\;(\mmin\,F\,(\suc\;\num{n}))  \\
\byval^{\vec{\sigma}}_\tau\;F\;\vec{X}\;\num{n}  & \reducesto & F \;\vec{X}\;\num{n} \;, 
                                                    \mbox{~~where $|\vec{X}|=|\vec{\sigma}|$}
\end{eqnarray*}
and with our repertoire of basic evaluation contexts augmented by
\[ \rec_\sigma\,X\,F\,[-] ~~~~~~~~~~ \mmin\,F\,[-] ~~~~~~~~~~ 
   \byval^{\vec{\sigma}}_\tau\,F\,\vec{X}\,[-]  \mbox{~~where $|\vec{X}|=|\vec{\sigma}|$} \;. \]

The constants $Y_\sigma$, $\rec_\sigma$ are familiar from $\PCF$ and System~$\sysT$ respectively,
whilst $\mmin$ is an inessential variant of the standard minimization operator $\mu$.
The operator $\while_\sigma$ is designed to capture that behaviour of a \textit{while} loop that 
manipulates data of type $\sigma$: 
the argument $C$ is the looping condition (with $\nat$ doing duty for the booleans, and 0 as true),
$X$ is the initial value of the data, and $F$ is the transformation applied to the data on each iteration.
The result returned by $\while_\sigma\,C\,F\,X$ is then the final value of the data when the loop terminates
(if it does).

The operator $\byval$ has a different character. 
The idea is that the evaluation of $\byval^{\vec{\sigma}}_\tau\;F\;\vec{X}\;N ~:\tau$ 
(where $|\vec{X}|=|\vec{\sigma}|$) will proceed by first trying to compute the value $\num{n}$ of $N$,
and if this succeeds, will then call $F\;\vec{X}$ `by value' on $\num{n}$.
This does essentially the same job as the operator $\byval$ of \cite[Section~7.1]{HOC},
which in our present notation would be written as $\byval^\epsilon_\nat$.
Indeed, for many purposes one could identify $\byval^{\vec{\sigma}}_\tau$ with
\[ \lambda f \vec{x} n \vec{y}.~\byval^\epsilon_\nat\, (\lambda n'.\,f \vec{x} n' \vec{y})\; n \;, \]
but there is a fine-grained difference in reduction behaviour which will matter for 
Proposition~\ref{PCF-faithful} below.

Our languages of interest are obtained by extending the definition of $\sysB$ as follows:
\begin{itemize}
\item For $\PCF$, we add the constant $Y_\sigma$ for each type $\sigma$.
\item For $\sysW$, we add the constants $\while_\sigma$.
\item For $\sysT$, we add the constants $\rec_\sigma$.
\end{itemize}
We shall also consider the further extensions $\sysT + \mmin$ and $\PCF + \byval$,
where we take the latter to include all $\byval^{\vec{\sigma}}_\tau$.

Note that in each case, the reduction relation is generated inductively from the specified basic reduction rules
together with the clause `if $M \reducesto M'$ then $E[M] \reducesto E[M']$',
where $E[-]$ ranges over the appropriately augmented set of basic evaluation contexts.
We thus obtain reduction relations $\reducesto_\PCF$, $\reducesto_\sysW$ etc.\ 
on the relevant sets of terms.
However, we may, without risk of ambiguity, write $\reducesto$ for the union of all these reduction relations,
noting that $\reducesto$ is still deterministic, 
and that if $M$ belongs to one of our languages $\LLL$ and $M \reducesto M'$ then $M'$ belongs to $\LLL$.

We write $\reducesto^+$ for the transitive closure of $\reducesto$, and $\reducesto^*$
for its reflexive-transitive closure.
It is easy to see that if $M$ is any closed term of type $\nat$,
then either $M \reducesto^* \num{n}$ for some unique $n \in \N$,
or the unique reduction path starting from $M$ is infinite;
in the latter case we say that $M$ \emph{diverges}.

The following fundamental fact will be useful.
It was first proved by Milner \cite{Milner-PCF} for $\PCF$, but extends readily to $\PCF+\byval$
(cf.\ \cite[Subsection~7.1.4]{HOC}).
Recall that two closed $\PCF+\byval$ terms $M,M':\sigma$ are \emph{observationally equivalent}
(written $M \obseq M'$)
if for every program context $C[-]:\nat$ of $\PCF+\byval$ (with a hole of type $\sigma$) 
and every $n \in \N$, we have
$C[M] \reducesto^* \num{n}$ iff $C[M'] \reducesto^* \num{n}$.

\begin{theorem}[Context lemma for $\PCF+\byval$]  \label{PCF-context-lemma}

(i) Suppose $M,M'$ are closed terms of type $\sigma_0 \arrow\cdots\arrow \sigma_{r-1} \arrow \tau$.
Then $M \obseq M'$ iff for all closed $N_0:\sigma_0$, \ldots, $N_{r-1}:\sigma_{r-1}$ we have
\[ M N_0 \ldots N_{r-1} ~\obseq~ M' N_0 \ldots N_{r-1} \;. \]

(ii) For closed $M,M': \sigma\times\tau$, we have $M \obseq M'$ iff $\fst\,M \obseq \fst\,M'$ and
$\snd\,M \obseq \snd\,M'$.

(iii) For closed $M,M':\nat$, we have $M \obseq M'$ iff $M,M'$ either both diverge or both evaluate
to the same numeral $\num{n}$. 
\end{theorem}

Let us also write $\equiv$ for the congruence on $\PCF+\byval$ terms generated by $\reducesto$
(i.e.\ the least equivalence relation containing $\reducesto$ and respected by all term contexts $C[-]$).
Clearly if $M \reducesto M'$ then $M \obseq M'$; hence also if $M \equiv M'$ then $M \obseq M'$.
In combination with the context lemma, this provides a powerful tool for establishing 
observational equivalences.

For any type $\sigma$, we write $\bot_\sigma$ for the `everywhere undefined' program
$Y_\sigma(\lambda x^\sigma.x)$.
It is not hard to see that if $M:\sigma$ admits an infinite reduction sequence then $M \obseq \bot_\sigma$.

At this point, we may note that the addition of $\byval$ does not fundamentally affect the expressive
power of $\PCF$, since as a simple application of the context lemma, we have
\[ \byval^{\vec{\sigma}}_\tau ~\obseq~ 
    \lambda f\vec{x}n\vec{y}.\;\ifzero\,n\,(f\vec{x}n\vec{y})\,(f\vec{x}n\vec{y}) \;. \]
This in turn implies that every $\PCF+\byval$ term is observationally equivalent to a $\PCF$ term, 
and also that it makes no difference to the relation $\obseq$ whether the
observing contexts $C[-]$ are drawn from $\PCF+\byval$ or just $\PCF$.
Even so, we shall treat $\byval$ as a separate language primitive rather than as a macro
for the above $\PCF$ term, since its evaluation behaviour is significantly different
(cf.\ Subsection~\ref{sec-NSPs} below).

We now show how both $\sysW$ and $\sysT+\mmin$ may be translated into $\PCF+\byval$.
To do this, we simply need to provide $\PCF+\byval$ programs of the appropriate types to
represent the constants $\while_\sigma$, $\rec_\sigma$, $\mmin$.
As a first attempt, one might consider natural implementations of these operations
along the following lines:
\begin{eqnarray*}
\mathit{While}_\sigma & = & \lambda c\,xf.\;Y_{\sigma\arrow\sigma} 
                                           (\lambda w.\,\ifzero\;(c\,x)\;(w(f\,x))\;x) \\ 
\mathit{Rec}_\sigma    & = & 
\lambda xf.\; Y_{\nat\arrow\sigma} (\lambda r.\lambda n.\,\ifzero\;n\;x\;(f(r(\pre\,n))(\pre\,n))) \\
\mathit{Min}  & = & 
\lambda f.\; Y_{\nat\arrow\nat} (\lambda m.\lambda n.\,\ifzero\;(f\,n)\;n\;(m\,(\suc\,n)))
\end{eqnarray*}
In fact, the above program $\mathit{While}_\sigma$ will serve our purpose as it stands,
but for $\mathit{Rec}_\sigma$ and $\mathit{Min}$ we shall need to resort to more complicated programs
that mimic the reduction behaviour of $\rec_\sigma$, $\mmin$
in a more precise way, using $\byval$ to impose a certain evaluation order.
We abbreviate the type of $\rec_\sigma$ as $\rho$, and the type of $\mmin$ as $\mu$;
we also write $\byval^+_\sigma$ for $\byval^{\sigma,(\sigma\arrow\nat\arrow\sigma)}_\sigma$.
\begin{eqnarray*}
\mathit{Rec}_\sigma    & = & 
\byval^+_\sigma ~ (Y_{\rho} \,(\lambda r^{\rho}. \\
& & ~~~\lambda xfn.\; \ifzero\;n\;x\;
      (\byval^\epsilon_\nat\;(\lambda n'.\,
      f \,((\byval^+_\sigma\,r)xfn')\;n')\,(\pre\;n)))) \\
\mathit{Min}  & = & 
\byval^{\nat\arrow\nat}_\nat ~ (Y_\mu \, (\lambda m^\mu.\; 
      \lambda fn.\; \ifzero\;(f\,n)\;n\;((\byval^{\nat\arrow\nat}_\nat\,m)\, f\, (\suc\,n))))
\end{eqnarray*}
We may then translate a term $M:\sigma$ of $\sysW$ or $\sysT+\mmin$ to a term $M^\circ:\sigma$
of $\PCF+\byval$ simply by replacing each occurrence of $\while_\sigma$, $\rec_\sigma$, $\mmin$
by the corresponding $\PCF+\byval$ program.
The following facts are routine to check by induction on the generation of $\reducesto$:

\begin{proposition}  \label{PCF-faithful}
(i) If $M \reducesto M'$ then $M^\circ \reducesto^+ {M'}^\circ$.

(ii) If $M^\circ \reducesto N$ then there is some $M'$ such that $M \reducesto M'$.

(iii) Hence $M \reducesto^* \num{n}$ iff $M^\circ \reducesto^* \num{n}$.
\QED
\end{proposition}

Next, we show that $\sysW$ and $\sysT+\mmin$ are also intertranslatable, though in a looser sense.
First, in either of these languages, it is an easy exercise to write a program
$\neq : \nat\arrow\nat\arrow\nat$ that implements inequality testing.
To assist readability, we shall use $\neq$ as an infix, and also allow ourselves some obvious 
pattern-matching notation for $\lambda$-abstractions on product types.
To translate from $\sysT+\mmin$ to $\sysW$, we use the $\sysW$ programs
\begin{eqnarray*}
\mathit{Rec}'_\sigma & = & \lambda xfn.~ \snd\;(
   \while_{\nat \times \sigma}~ 
   (\lambda \ang{n',x'}.\; n' \neq n)~ \\
& & ~~~~~~~~~~~~~~~   \ang{0,x}~~ (\lambda \ang{n',x'}.\; \ang{\suc\,n', fx'n'})
   ) \\
\mathit{Min}' & = & \lambda fn.\;\while_\nat~ (\lambda n'.\; f n' \neq 0)~n~\suc 
\end{eqnarray*}
We may then translate a $\sysT+\mmin$ term $M$ to a $\sysW$ term $M^\dag$
simply by replacing each occurrence of $\rec_\sigma$ $\mmin$ 
by $\mathit{Rec}'_\sigma$, $\mathit{Min}'$ respectively.
However, it will \emph{not} in general be the case for this translation that if $M \reducesto M'$ then
$M^\dag \reducesto^+ {M'}^\dag$: the operational behaviour of $M$ and $M^\dag$ at an intensional level
may be quite different. Nevertheless, we can show that the translation is faithful in the sense that
$M$ and $M^\dag$ are \emph{observationally equivalent} when both are transported to $\PCF$:

\begin{proposition}  \label{W-faithful}
(i) $(\mathit{Rec}'_\sigma)^\circ \obseq \mathit{Rec}_\sigma$
and $(\mathit{Min}')^\circ \obseq \mathit{Min}$ as $\PCF$ terms.

(ii) For any closed term $M$ of $\sysT+\mmin$, we have $(M^\dag)^\circ \obseq M^\circ$.

(iii) For any closed $M:\nat$ in $\sysT+\mmin$, we have
$M \reducesto^* n$ iff $M^\dag \reducesto^* n$.
\end{proposition}

\noindent \proofsk
(i) For $\mathit{Rec}_\sigma$, by Theorem~\ref{PCF-context-lemma} it suffices to show that for any 
closed PCF terms $X:\sigma$, $F:\sigma\arrow\nat\arrow\sigma$ and $N:\nat$, we have
\[ (\mathit{Rec}'_\sigma)^\circ\;X\;F\;N ~\obseq~ \mathit{Rec}_\sigma\;X\;F\;N  \;. \]
But this is routinely verified: if $N$ diverges then both sides admit infinite reduction sequences and so
are observationally equivalent to $\bot_\sigma$;
whilst if $N \reducesto^* n$ then an easy induction on $n$ shows that
\[ (\mathit{Rec}'_\sigma)^\circ\;X\;F\;n ~\equiv~ 
   F\,( \cdots (F\,(F\,X\,\num{0})\,\num{1}) \cdots)\,\num{n-1}
   ~\equiv~ \mathit{Rec}_\sigma\;X\;F\;n \;. \]
A similar approach works for $\mathit{Min}$.

(ii) follows immediately, since $(M^\dag)^\circ$ may be obtained from $M^\circ$ by
replacing certain occurrences of $\mathit{Rec}_\sigma, \mathit{Min}$ by 
$(\mathit{Rec}'_\sigma)^\circ$, $(\mathit{Min}')^\circ$.

(iii) is now immediate from (ii) and Proposition~\ref{PCF-faithful}(iii). 
\QED

\vspace*{1.5ex}
To translate from $\sysW$ to $\sysT+\mmin$, we may define
\begin{eqnarray*}
\mathit{While}'_\sigma & = & \lambda cxf.~ 
   \Rec_\sigma~x~(\lambda x'n.\,fx')~ \\
& & ~~~~~~~~~~~~~~~  (\mmin\;(\lambda n.\; \Rec_\sigma\;x\;(\lambda x'n.\,fx') \neq 0)\;0)
\end{eqnarray*}
and translate a $\sysW$ term $M$ to a $\sysT+\mmin$ term $M^\ddag$ by replacing each
$\while_\sigma$ with $\mathit{While}'$.
Once again, the operational behaviour of $M^\ddag$ is in general quite different from that of $M$:
indeed, this translation is grossly inefficient from a practical point of view, since typically some 
subcomputations will be repeated several times over.
Nonetheless, if we consider programs only up to observational equivalence, the translation is still
faithful in the way that we require:

\begin{proposition}  \label{Tmin-faithful}
(i) $(\mathit{While}'_\sigma)^\circ \obseq \mathit{While}_\sigma$.

(ii) For any closed term $M$ of $\sysW$, we have $(M^\ddag)^\circ \obseq M^\circ$.

(iii) For any closed $M:\nat$ in $\sysW$, we have
$M \reducesto^* n$ iff $M^\ddag \reducesto^* n$.
\end{proposition}

\noindent \proofsk
Closely analogous to Proposition~\ref{W-faithful}.
For (i), we show that by induction on $n$ that if $C(F^n(X)) \reducesto \num{0}$ whereas
$C(F^i(X)) \reducesto \num{m_i} \neq \num{0}$ for each $i<n$, then
\[ (\mathit{While}'_\sigma)^\circ\;C\;F\;X ~\equiv~ F^n(X)
   ~\equiv~ \mathit{While}_\sigma\;C\;F\;X \;. \]
Furthermore, we show that if there is no $n$ with this property, then
$ (\mathit{While}'_\sigma)^\circ\;C\;F\;X$ and $\mathit{While}_\sigma\;C\;F\;X$ both admit
infinite reduction sequences, so that both are observationally equivalent to $\bot_\sigma$.
\QED

\vspace*{1.5ex}
In summary, we have shown that $\sysT+\mmin$ and $\sysW$ are equally expressive as sublanguages
of $\PCF$, in the sense that a closed PCF term $M$ is observationally equivalent to (the image of)
a $\sysT+\mmin$ term iff it is observationally equivalent to a $\sysW$ term.
As already noted, the language $\sysW$ appears to embody a natural general principle of iteration,
suggesting that this level of expressive power is a natural one to consider.

Although not formally necessary for this paper, it is also worth observing that the above equivalence works 
level-by-level. 
For each $k \geq 0$, let us define sublanguages $\PCF_k$, $\sysT_k+\mmin$, $\sysW_k$
of $\PCF$, $\sysT+\mmin$, $\sysW$ by admitting (respectively) the constants
$Y_\sigma$, $\rec_\sigma$, $\while_\sigma$ only for types $\sigma$ of level $\leq k$.
Then the translations $-^\dag, -^\ddag$ clearly restrict to
translations between $\sysT_k + \mmin$ and $W_k$, so that $\sysT_k + \mmin$ and $W_k$ are
equally expressive as sublanguages of $\PCF$.%
\footnote{This hierarchy turns out to be strict, as will be shown in a forthcoming paper \cite{Sublangs-PCF}.
The strictness of the hierarchy of languages $\PCF_k$ is established in \cite{Y-hierarchy}.}
In fact, for $k \geq 1$, we can even regard them as sublanguages of $\PCF_k$,
since it is an easy exercise to replace our `precise' translation $-^\circ$ for
$\sysT_k + \mmin$ or $W_k$
by a translation $-^\bullet$ up to observational equivalence that requires only $\PCF_k$.
This does not work for $k=0$, however,
since the implementation of $\rec_\nat$ requires at least $Y_{\nat\arrow\nat}$.

\subsubsection{Weaker languages}  \label{sec-weaker-langs}

A few weaker languages will play more minor roles in the paper.
We introduce them here, suppressing formal justifications of points mentioned
merely for the sake of orientation.

To motivate these languages, 
we note that even the primitive recursor $\rec_\nat$ of $\sysT_0$ works in a `lazy' way:
it is possible for the value of $\rec_\nat\,X\,F\,\num{n+1}$ to be defined even if 
that of $\rec_\nat\,X\,F\,\num{n}$ is not, for example if $n=0$, $X = \bot_\nat$ and $F = \lambda xn.0$.
This contrasts with the `strict' behaviour of Kleene's original version of primitive recursion, in which
the value of $\rec_\nat\,X\,F$ at $\num{n}$ is obtained by successively computing its values
at $\num{0},\num{1},\ldots,\num{n-1}$, all of which must be defined.
(Of course, the distinction is not very visible in purely total settings such as $\Set$ or $\Ct$.)

%


We may capture this stricter behaviour with the help of 
$\byval^\epsilon_\nat: (\nat\arrow\nat)\arrow\nat\arrow\nat$, which we shall here write as 
$\byval_{[\nat]}$.
From this, we may inductively define an operator
\[ \byval_{[\sigma]} ~:~ (\sigma\arrow\sigma)\arrow\sigma\arrow\sigma \]
for each level 0 type $\sigma$ by:
\[  \byval_{[\sigma\times\tau]} ~=~ 
     \lambda f.\; \lambda x^{\sigma\times\tau}.\; 
     \byval_{[\sigma]}\;(\lambda y^\sigma.\, \byval_{[\tau]} (\lambda z^\tau.\,f \ang{y,z})\,(\snd\,x))\;(\fst\,x) \;. \]
We may now use these operators to define `strict' versions of $\rec_\sigma$ and $\while_\sigma$ for any type
$\sigma$ of level $0$. We do this by introducing constants
$\rec_\sigma^\str$ and $\while_\sigma^\str$ of the same types as $\rec_\sigma$, $\while_\sigma$, 
with reduction rules
\begin{eqnarray*}
\rec_\sigma^\str\;X\;F\;\num{0} & \reducesto & X \\
\rec_\sigma^\str\;X\;F\;\num{n+1} & \reducesto & \byval_{[\sigma]}\;(\lambda m.\,F\,m\,\num{n})\;(\rec_\sigma^\str\,X\,F\,\num{n}) \\
\while_\sigma^\str\;C\;X\;F & \reducesto & 
  \byval_{[\sigma]}\;(\lambda x.\,\ifzero\;(Cx)\;(\while_\sigma\;C\;(Fx)\;F) \; x) \; X 
\end{eqnarray*}
We also add $\rec_\sigma^\str\,X\,F\,[-]$ as a basic evaluation context.
By using these in place of their non-strict counterparts, and also including $\byval_{[\nat]}$
so that the operators $\byval_{[\sigma]}$ are available, we obtain languages $\sysT_0^\str$,
$\sysT_0^\str + \mmin$ and $W_0^\str$.

Our earlier PCF programs $\mathit{Rec}_\sigma$, $\mathit{While}_\sigma$ may be readily adapted 
to yield faithful translations of these languages into $\PCF+\byval$.
It can also be checked that $\sysT_0^\str + \mmin$ and $W_0^\str$ are
intertranslatable in the same way as $\sysT_0 + \mmin$ and $W_0$, and so are equi-expressive as
sublanguages of $\PCF$.
Finally, we can regard the strict versions as sublanguages of the lazy ones up to observational equivalence,%
\footnote{It is shown in \cite[Theorem~6.3.23]{HOC} that $\rec_0^\str$ is strictly weaker than $\rec_0$
(note that $\sysT_0 + \mmin$ is in essence the language known as $\Klex^\smallmin$ in \cite{HOC}).
What is perhaps more surprising is that $\sysT_0 + \mmin$ defines more \emph{total} functionals
at third order than does $\sysT_0^\str + \mmin$, despite the fact that $\rec_0$, $\rec_0^\str$ have the
same behaviour on all total arguments \cite{Sublangs-PCF}.}
since for example
\[ \rec_\sigma^\str ~\obseq~
   \lambda xfn.\; \byval_{[\sigma]}\; (\lambda x'.\; \rec_\sigma\,x\,(\lambda ym.\,\byval(\lambda y'.fy'n)\,y)\,n) \;x \;.
\]
and it is easy to supply terms of B observationally equivalent to each $\byval_{[\sigma]}$.

It is natural to think of $\sysW_0^\str$ as the language for `everyday' iterative computations on ground data.
It is easy to check that at type level 1, $\sysT_0^\str + \mmin$ and $\sysW_0^\str$ 
define all Turing computable functions, 
and indeed that (respectively) $\rec_\nat^\str$ and $\while_{\nat\times\nat}^\str$ 
are sufficient for this purpose.

The weakest language of all that we shall consider is $\sysT_0^\str$.
This will play an ancillary as a language for a rudimentary class of 
`non-controversially total' functionals present in all settings of interest,
enabling us to formulate our main theorem in a robust and portable form.
In all total type structures of interest, the $\sysT_0^\str$ definable
functionals will clearly coincide with those given by Kleene's S1--S8; 
at type level 1 these are just the usual primitive recursive functions.
However, a somewhat subtle point is that for typical interpretations 
in \emph{partial} type structures, $\sysT_0^\str$ will be slightly stronger
than the usual formulations of S1--S8, since the constant $\ifzero$ will give us the
power of \emph{strong definition by cases} which is not achievable via S1--S8 alone.
This point will be significant in Section~\ref{sec-main-theorem}, where we shall frequently claim that
certain elements of the model $\SP^0$ are $\sysT_0^\str$ definable;
the reader should bear in mind here that strong definitions by cases are permitted.%
\footnote{In \cite{HOC}, this issue was addressed by introducing a specially defined class
$\SP^{0,\prim+}$ of \emph{strongly total} elements of $\SP^0$.}


\subsubsection{Elimination of product types}

So far, we have worked with languages with product types in order to manifest
the equivalence of $\sysT+\min$ and $\sysW$ (and various restrictions thereof) in a perspicuous way. 
However, since the bar recursors that are the subject of our main theorem have a type not involving products,
it will be sufficient from here on to work with types without $\times$, and it will simplify the 
presentation of nested sequential procedures (in the next subsection) to do so.
From the point of view of expressivity, nothing of significance is lost by dispensing with product types,
in view of the following proposition. 
Here we say a type $\sigma$ is \emph{$\times$-free} if it does not involve products,
and a term $M$ is \emph{$\times$-free} if the types of $M$ and all its subterms are $\times$-free.
In particular, a $\times$-free term $M$ may involve operators $Y_\sigma$, $\rec_\sigma$, $\while_\sigma$
only for $\times$-free $\sigma$.
We shall write $\id_\sigma$ for $\lambda x^\sigma.x$, and $g \circ f$ for $\lambda x.g(f\,x)$.

\begin{proposition}  \label{dont-need-products}
Suppose $\sigma$ is $\times$-free.
Then any closed term $M:\sigma$ of $\PCF$ (resp.\ $\sysT+\mmin$, $\sysW$) is observationally equivalent 
to a $\times$-free term of the same language.
\end{proposition}

\noindent \proofsk
This will be clear from familiar facts regarding the embeddability of arbitrary types in $\times$-free ones,
as covered in detail in \cite[Section~4.2]{HOC}.
More specifically, we may specify, for each type $\sigma$, a $\times$-free type $\widehat{\sigma}$
such that $\sigma$ is a $\sysT_0^\str$ definable retract of $\widehat{\sigma}$ 
up to observational equivalence: that is, there are closed B terms 
$\enc_\sigma: \sigma \arrow \widehat{\sigma}$ and $\dec_\sigma: \widehat{\sigma} \arrow \sigma$
such that $\lambda x^\sigma.\,\dec_\sigma(\enc_\sigma\,x) \;\obseq\; \id_\sigma$.
Moreover, we may choose these data in such a way that 
\begin{itemize}
\item $\widehat{\nat} = \nat$ and $\widehat{\sigma\arrow\tau} = \widehat{\sigma\arrow\tau}$,
and moreover we have $\enc_\nat = \dec_\nat = \id_\nat$,
$\enc_{\sigma\arrow\tau} = \lambda f.\,\enc_\tau \circ f \circ \dec_\sigma$, and
$\dec_{\sigma\arrow\tau} = \lambda g.\,\dec_\tau \circ g \circ \enc_\sigma$
(these facts imply that for all $\times$-free $\sigma$ we have $\widehat{\sigma}=\sigma$ and 
$\enc_\sigma \obseq \dec_\sigma \obseq \id_\sigma$),
\item pairing and projections are represented relative to this encoding by $\times$-free programs
$\mathit{Pair}: \widehat{\sigma}\arrow\widehat{\tau}\arrow\widehat{\sigma\times\tau}$,
$\mathit{Fst}: \widehat{\sigma\times\tau} \arrow \widehat{\sigma}$, 
$\mathit{Snd}: \widehat{\sigma\times\tau} \arrow \widehat{\tau}$,
\item for any $\sigma$ we have $Y_\sigma \obseq \dec_\sigma(Y_{\widehat{\sigma}})$,
and similarly for $\rec_\sigma$ and $\while_\sigma$.
\end{itemize}
Using these facts, it is easy to construct a compositional translation assigning 
to each term $M:\sigma$ (with free variables $x_i:\sigma_i$) 
a $\times$-free term $\widehat{M} : \widehat{\sigma}$ 
(with free variables $\widehat{x_i}:\widehat{\sigma_i}$)
such that $M \obseq \dec_\sigma (\widehat{M}[\widehat{\vec{x}} \mapsto \enc(\vec{x})])$
(we omit the uninteresting details).
In particular, for closed $M$ of $\times$-free type $\sigma$, this yields $M \obseq \widehat{M}$,
which achieves our purpose.
\QED

\vspace*{1.5ex}
From here onwards, we shall therefore use the labels $\PCF$, $\sysT+\mmin$, $\sysW$, etc.\
to refer to the $\times$-free versions of these languages, and shall only refer to types generated
from $\nat$ via $\arrow$.

\subsection{Nested sequential procedures}  \label{sec-NSPs}

Next, we summarize the necessary elements of the theory of \emph{nested sequential procedures} (NSPs)
also known as \emph{PCF B\"ohm trees},%
\footnote{The term `nested sequential procedure' was adopted in \cite{HOC} as a neutral label for 
a notion that is of equal relevance to both PCF and Kleene computability.}
relying on \cite{HOC} for further details and for the relevant proofs.
Although we shall provide enough of the formal details to support what we wish to do,
a working intuition for NSPs is perhaps more easily acquired by looking at examples. 
The reader may therefore wish to look at the examples appearing from Definition~\ref{PCF-interp-def}
onwards in conjunction with the following definitions.

As explained in Subsection~3.2.5 and Section~6.1 of \cite{HOC},
\emph{nested sequential procedures} (or \emph{NSPs}) are infinitary terms
generated by the following grammar, construed coinductively:
\begin{eqnarray*}
\mbox{\it Procedures:}~~~~
p,q & ::= & \lambda x_0 \ldots x_{r-1}.\,e \\
\mbox{\it Expressions:}~~~~
d,e & ::= & \bot ~\mid~ n ~\mid~ \ccase~a~\,\oof~ 
            (i \darrow e_i \mid i \in \N) \\
\mbox{\it Applications:}~~~~~~~
a   & ::= & x\, q_0 \ldots q_{r-1}
\end{eqnarray*}
Informally, an NSP captures the possible behaviours of a (sequential) program with inputs bound
to the formal parameters $x_j$, which may themselves be of function type. 
Such a program may simply diverge ($\bot$), or return a value $n$,
or apply one of its inputs $x_j$ to some arguments---the subsequent behaviour of the program
may depend on the numerical result $i$ of this call.
Here the arguments to which $x_j$ is applied are themselves specified via NSPs $q_0,\ldots,q_{r-1}$
(which may also involve calls to $x_j$).
In this way, NSPs should be seen as syntax trees which may be infinitely deep as well as infinitely broad.

We use $t$ as a meta-variable ranging over all three kinds of NSP terms.
We shall often use vector notation 
$\myvec{x}, \myvec{q}$ for finite sequences
$x_0 \ldots x_{r-1}$ and $q_0 \ldots q_{r-1}$.
Note that such sequences may be empty, so that for instance we have
procedures of the form $\lambda.e$.
As a notational concession, we will sometimes denote the application of a variable $x$
to an empty list of arguments by $x()$.
If $p = \lambda \vec{y}.e$, we will also allow the notation $\lambda x.p$ to mean $\lambda x\vec{y}.e$.
The notions of free variable and (infinitary) $\alpha$-equivalence
are defined in the expected way, and we shall work with terms
only up to $\alpha$-equivalence.

If variables $x$ are considered as annotated with simple types $\sigma$ (as indicated by writing $x^\sigma$),
there is an evident notion of a \emph{well-typed} procedure term $p: \sigma$, where $\sigma \in \Tsf$. 
Specifically, within any term $t$, occurrences of procedures $\lambda \vec{x}.e$ (of any type), 
applications $x \vec{q}$ (of the ground type $\nat$) and expressions $e$ (of ground type) 
have types that are related to the types of their constituents and of variables as usual in typed 
$\lambda$-calculus extended by case expressions of type $\nat$. 
We omit the formal definition here since everything works as expected; 
for a more precise formulation see [15, Section 6.1.1]. 

For each type $\sigma$, we let $\SP(\sigma)$ be the set of well-typed procedures of type $\sigma$, 
and $\SP^0(\sigma)$ for the set of \emph{closed} such procedures; note that $\SP^0(\nat) \iso \N_\bot$.
As a notational liberty, we will sometimes write the procedures $\lambda.n, \lambda.\bot \in \SP^0(\nat)$
simply as $n,\bot$.

Note that in the language of NSPs, one cannot directly write e.g.\ $f^{\sigma\arrow\nat}x^\sigma$,
since the variable $x^\sigma$ is not formally a procedure.
However, we may obtain a procedure corresponding to $x^\sigma$ via
\emph{hereditary $\eta$-expansion}:
if $\sigma = \sigma_0 \arrow\cdots\arrow \sigma_{r-1} \arrow \nat$,
then we define a procedure $x^{\sigma\eta}$ inductively on types by
\[ x^{\sigma\eta} ~=~ \lambda z_0^{\sigma_0} \cdots z_{r-1}^{\sigma_{r-1}}.\;
   \caseof{x z_0^{\sigma_0 \eta} \cdots z_{r-1}^{\sigma_{r-1}\eta}}{i \darrow i} \;. \]
For example, the procedure corresponding to the identity on type $\sigma$ may now be written as
$\lambda x^\sigma. x^{\sigma\eta}$.
Note that $x^{\nat\eta} = \lambda.\,\caseof{x()}{i \darrow i}$.

In order to perform computation with NSPs, and in particular to define the \emph{application} of a procedure
$p$ to an argument list $\vec{q}$, we shall use an extended calculus of \emph{meta-terms}, within which
the terms as defined above will play the role of normal forms.
Meta-terms are generated by the following infinitary grammar, again construed coinductively:
\begin{eqnarray*}
\mbox{\it Meta-procedures:}~~~~
P,Q & ::= & \lambda \vec{x}.\,E \\
\mbox{\it Meta-expressions:}~~~~
D,E & ::= & \bot ~\mid~ n ~\mid~ \ccase~G~\,\oof~ 
            (i \darrow E_i \mid i \in \N) \\
\mbox{\it Ground meta-terms:}~~~~~~~
G   & ::= & E ~\mid~ x\, \vec{Q} ~\mid~ P \vec{Q}
\end{eqnarray*}
Again, our meta-terms will be subject to the evident typing rules which work as expected.
Unlike terms, meta-terms are amenable to a notion of (infinitary) substitution:
if $T$ is a meta-term and $Q_0,\ldots,Q_{r-1}$ are meta-procedures whose types match those of
$x_0,\ldots,x_{r-1}$ respectively, we have the evident meta-term $T[\vec{x} \mapsto \vec{Q}]$.

We equip our meta-terms with a \emph{head reduction} $\reducesto_h$ generated as follows:
\begin{itemize}
\item $(\lambda \vec{x}.E) \vec{Q} \reducesto_h E[\vec{x} \mapsto \vec{Q}]$ ~~($\beta$-rule).
\item $\caseof{\bot}{i \darrow E_i} \reducesto_h \bot$.
\item $\caseof{n}{i \darrow E_i} \reducesto_h E_n$.
\item $\caseof{(\caseof{G}{i \darrow E_i})}{j \darrow F_j} \reducesto_h$ \\
         \hspace*{2ex}  $\caseof{G}{i \darrow \caseof{E_i}{j \darrow F_j}} \).
\item If $G \reducesto_h G'$ and $G$ is not a $\ccase$ meta-term, then
        \[ \caseof{G}{i \darrow E_i} \reducesto_h \caseof{G'}{i \darrow E_i} \;. \]
\item If $E \reducesto_h E'$ then $\lambda \vec{x}.E \reducesto_h \lambda \vec{x}.E'$.
\end{itemize}
We write $\reducesto_h^*$ for the reflexive-transitive closure of $\reducesto_h$.
We call a meta-term a \emph{head normal form} if it cannot be further reduced using $\reducesto_h$.
The possible shapes of meta-terms in head normal form are 
$\bot$, $n$, $\caseof{y \vec{Q}}{i \darrow E_i}$ and $y \vec{Q}$, 
the first three optionally prefixed by $\lambda \vec{x}$.

We may now see how an arbitrary meta-term $T$ may be evaluated to a normal form $t$, by a 
process analogous to the computation of B\"ohm trees in untyped $\lambda$-calculus.
For the present paper a somewhat informal description will suffice;
for a more formal treatment we refer to \cite[Section~6.1]{HOC}.
First, we attempt to reduce $T$ to head normal form by repeatedly applying head reductions.
If a head normal form is never reached, the normal form is $\bot$ 
(possibly prefixed by some $\lambda \vec{x}$ appropriate to the type).
If the head normal form is $\bot$ or $n$ (possibly prefixed by $\lambda \vec{x}$), 
then this is the normal form of $T$. 
If the head normal form is $\caseof{y \vec{Q}}{i \darrow E_i}$, then we recursively evaluate 
the $Q_j$ and $E_i$ to normal forms $q_j, e_i$ by the same method, and take 
$t = \caseof{y \vec{q}}{i \darrow e_i}$; likewise for $\ccase$ expressions prefixed by a $\lambda$,
and for meta-terms $y \vec{Q}$.
Since the resulting term $t$ may be infinitely deep, this evaluation is in general an infinitary process
in the course of which $t$ crystallizes out, although any required finite portion of $t$ may be computed by
just finitely many reductions.

If $p = \lambda x_0 \cdots x_r. e \in \SP(\sigma\arrow\tau)$ and $q \in \SP(\sigma)$,
we define the \emph{application}
$p \cdot q \in \SP(\tau)$ to be the normal form of the meta-procedure
$\lambda x_1 \cdots x_r.\, e[x_0 \mapsto q]$.
This makes the sets $\SP(\sigma)$ into a total applicative structure $\SP$,
and the sets $\SP^0(\sigma)$ of closed procedures into a total applicative structure $\SP^0$.


We write $\sqsubseteq$ for the syntactic ordering on each $\SP(\sigma)$, so that $p \sqsubseteq p'$
if $p$ is obtained from $p'$ by replacing certain subterms (perhaps infinitely many) by $\bot$.
It is not hard to check that each $\SP^0(\sigma)$ is a DCPO with this ordering, and that 
application is monotone and continuous with respect to this structure.

We also have an \emph{extensional preorder} $\preceq$ on each $\SP^0(\sigma)$ defined as follows:
if $p,p' \in \SP^0(\sigma)$ where $\sigma = \sigma_0 \arrow \cdots \arrow \sigma_{t-1} \arrow \nat$,
then
\[ p \preceq p' \mbox{~~iff~~} \forall q_0,\ldots,q_{t-1}.\forall n.\;
   (p \cdot q_0 \cdot \ldots \cdot q_{t-1} = \lambda.n) ~\Rightarrow~
   (p' \cdot q_0 \cdot \ldots \cdot q_{t-1} = \lambda.n) \]
The following useful fact is established in Subsection~6.1.4 of \cite{HOC}:

\begin{theorem}[NSP context lemma]   \label{NSP-context-lemma}
If $p \preceq p' \in \SP^0(\sigma)$, then for all $r \in \SP^0(\sigma\arrow\nat)$ we have
$r \cdot p \sqsubseteq r \cdot p'$.
\end{theorem}

We now have everything we need to give an interpretation of simply typed $\lambda$-calculus in
$\SP$: a variable $x^\sigma$ is interpreted by $x^{\sigma\eta}$, application is interpreted by $\cdot$, 
$\lambda$-abstraction is interpreted by itself, and we may also add a constant $p$ for each $p \in \SP^0$
(interpreted by itself).
The following non-trivial theorem, proved in \cite[Section~6.1]{HOC}, ensures that many familiar kinds of 
reasoning work smoothly for NSPs:

\begin{theorem}
$\SP^0$ is a \emph{typed $\lambda\eta$-algebra}: 
that is, if $U,V:\sigma$ are closed simply typed $\lambda$-terms
with constants drawn from $\SP^0$ and $U =_\beta V$, then $U,V$ denote the same element of 
$\SP^0(\sigma)$ under the above interpretation.
\end{theorem}

One of the mathematically interesting aspects of $\SP^0$ is the existence of several well-behaved 
\emph{substructures} corresponding to more restricted flavours of computation.
The two substructures of relevance to this paper are defined as follows:

\begin{definition}
(i) The \emph{application tree} of an NSP term $t$ is simply the tree of all occurrences of applications
$x \vec{q}$ within $t$, ordered by subterm inclusion.

(ii) An NSP term $t$ is \emph{left-well-founded} (LWF) if its application tree is well-founded.

(iii) A term $t$ is \emph{left-bounded} if its application tree is of some finite depth $d$.
\end{definition}

The following facts are proved in \cite[Section~6.3]{HOC}:

\begin{theorem}  \label{substructures-thm}
(i) LWF procedures are closed under application.
Moreover, the substructure $\SP^{0,\lwf}$ of $\SP^0$ consisting of LWF procedures is
a sub-$\lambda\eta$-algebra of $\SP^0$.

(ii) Left-bounded procedures are closed under application, and the corresponding substructure
$\SP^{0,\lbd}$ is also a sub-$\lambda\eta$-algebra of $\SP^0$.
\end{theorem}

Next, we indicate how NSPs provide us with a good model for the behaviour of terms of $\PCF+\byval$.

\begin{definition}  \label{PCF-interp-def}
To any $\PCF+\byval$ term $M:\sigma$ (possibly with free variables) we may associate a procedure
$\sem{M} \in \SP(\sigma)$ (with the same or fewer free variables) in a compositional way,
by recursion on the term structure of $M$:
\begin{itemize}
\item $\sem{x^\sigma} = x^{\sigma\eta}$.
\item $\sem{\lambda x.M} = \lambda x.\sem{M}$.
\item $\sem{MN} = \sem{M} \cdot \sem{N}$
\item $\sem{\suc} = \lambda x^\nat.\,\caseof{x()}{0 \darrow 1 \,\mid\, 1 \darrow 2 \,\mid\, 2 \darrow 3 \,\mid\, \cdots}$.
\item $\sem{\pre} = \lambda x^\nat.\,\caseof{x()}{0 \darrow 0 \,\mid\, 1 \darrow 0 \,\mid\, 2 \darrow 1 \,\mid\, \cdots}$.
\item $\sem{\ifzero} = \lambda x^\nat y^\nat z^\nat.\;\ccase~{x()}~\oof~(0 \darrow \caseof{y()}{j \darrow j}$ \\ 
\hspace*{31ex} $\mid~ i+1 \darrow \caseof{z()}{j \darrow j})$.
\item $\sem{\byval^{\vec{\sigma}}_\tau} = \lambda f^{\vec{\sigma}\arrow\nat\arrow\tau} 
\vec{x} n \vec{y}.\;\ccase~{n()}~\oof~(0 \darrow \caseof{f \vec{x}^{\,\eta} (\lambda.0) \vec{y}^{\,\eta}}{j \darrow j}$ \\
\hspace*{37ex} $\mid~ 1 \darrow \caseof{f \vec{x}^{\,\eta} (\lambda.1) \vec{y}^{\,\eta}}{j \darrow j}$ \\
\hspace*{37ex} $\mid~ \cdots~ )$.
\item If $\sigma = \sigma_0 \arrow \cdots \sigma_{r-1} \arrow \nat$, then $\sem{Y_\sigma}$ is
the NSP depicted below:
\end{itemize}
\vspace*{-2.5ex}
\begin{center}
\includegraphics[scale=.60]{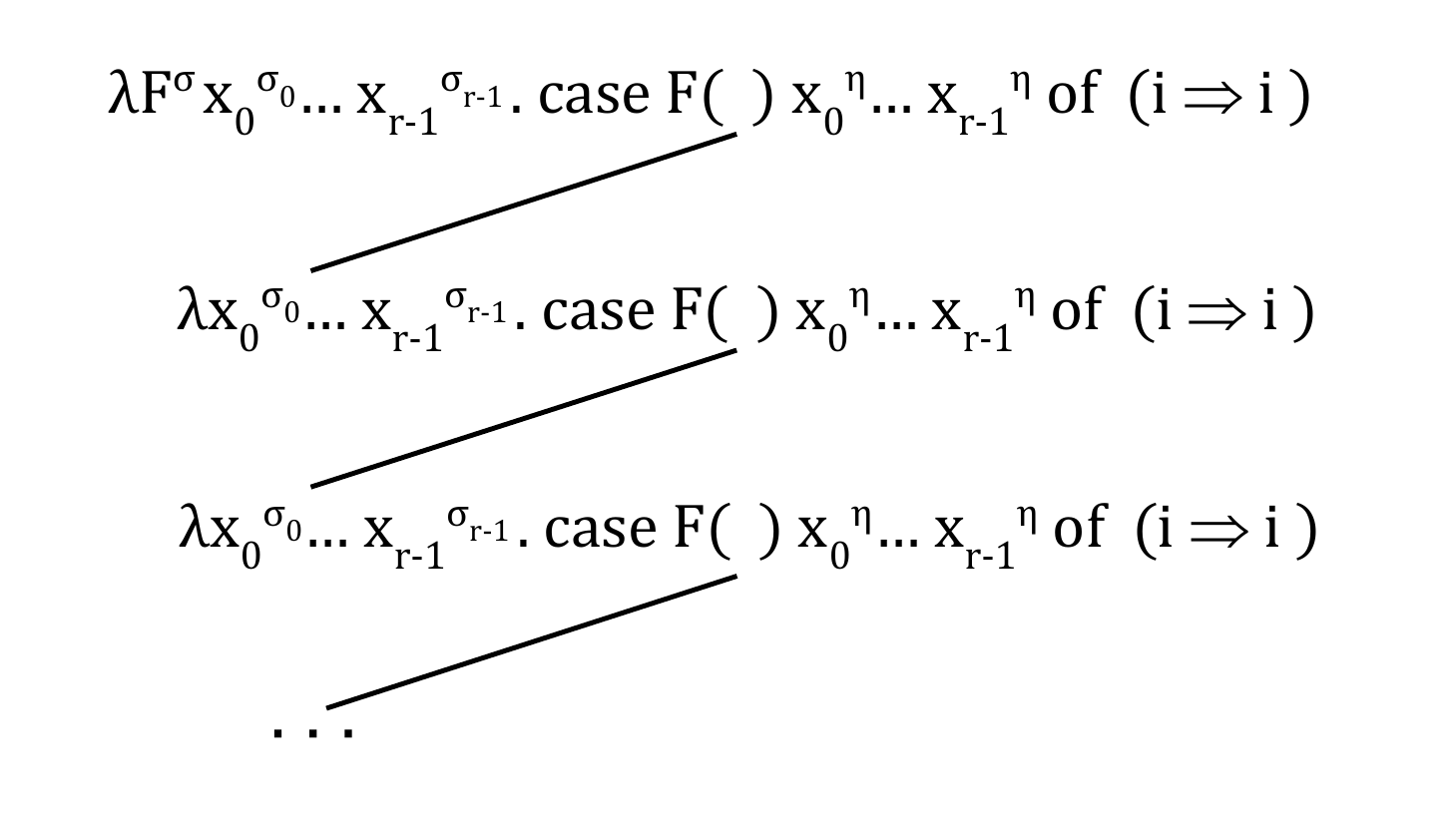}
\end{center}
\end{definition}

\vspace*{-2.0ex}
The NSP for $Y_\sigma$ is the archetypal example of a non-LWF procedure:
the nested sequence of application subterms $F(\cdots)$ never bottoms out.

The following theorem, proved in \cite[Subsection~7.1.3]{HOC}, confirms that this interpretation is faithful
to the behaviour of $\PCF+\byval$ programs:
\begin{theorem}[Adequacy of NSP model]
If $M$ is any closed $\PCF+\byval$ term of type $\nat$, then 
$M \reducesto^* n$ iff $\sem{M} = \lambda.n$, and $M$ diverges iff $\sem{M} = \lambda.\bot$.
\end{theorem}
It is also the case that every \emph{computable} element of any $\SP^0(\sigma)$ is denotable by a
closed $\PCF+\byval$ term of type $\sigma$ (see \cite[Subsection~7.1.5]{HOC}), 
though we shall not need this fact in this paper.

We also obtain interpretations of $\sysT+\mmin$ and $\sysW$ in $\SP^0$, induced by the translations
of these languages into $\PCF$ as described in Subsection~\ref{sec-languages}.
Applying the definition of $\sem{-}$ above to the $\PCF$ programs 
$\mathit{Rec}_\sigma$, $\mathit{Min}$, $\mathit{While}_\sigma$, 
we may thus obtain the appropriate NSPs for the constants $\rec_\sigma$, $\mmin$, $\while_\sigma$
respectively. To avoid clutter, we allow an application term $a$ to stand for the expression
$\caseof{a}{i \darrow i}$. Where a case branch label involves a metavariable $i$ or $j$,
there is intended to be a subtree of the form displayed for each $i,j \in \N$.

\vspace*{-2.0ex}
\begin{center}
\includegraphics[scale=.60]{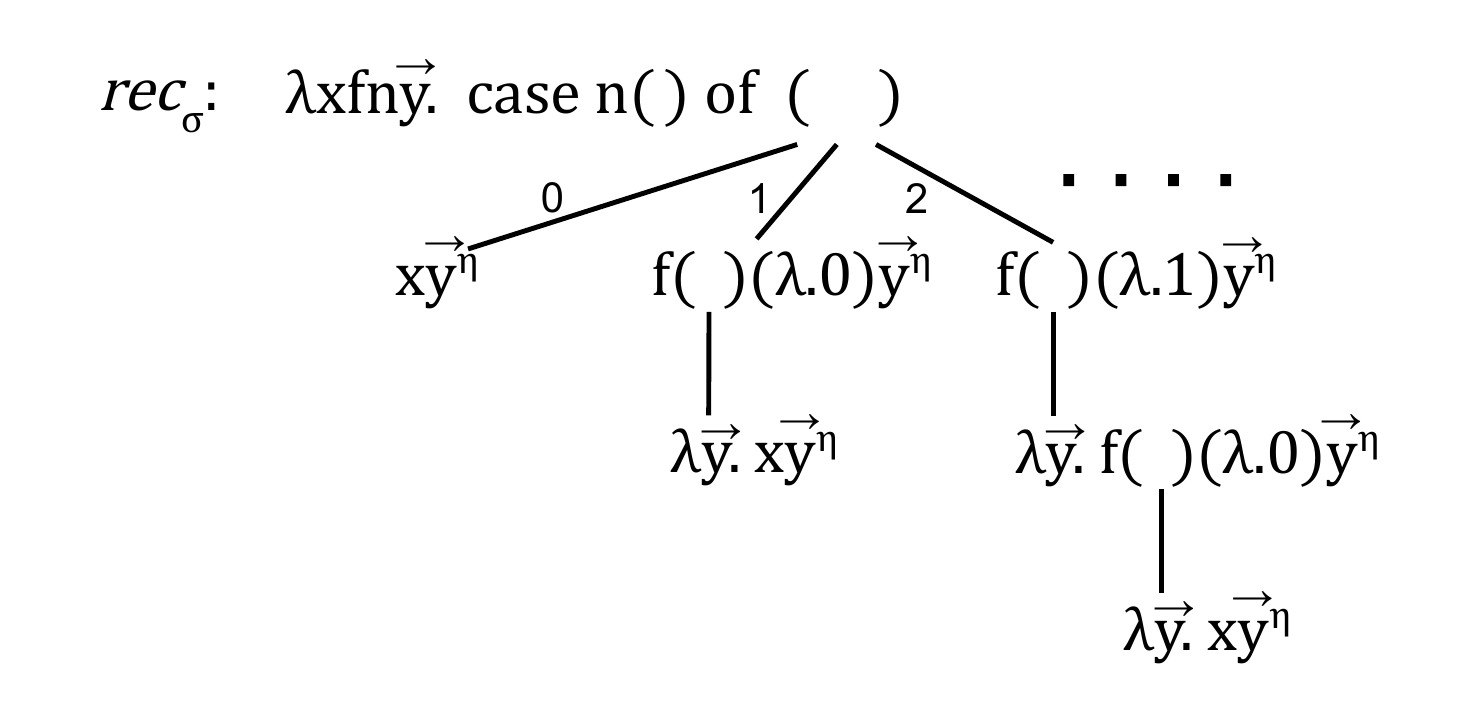}
\vspace*{-1.5ex}
\includegraphics[scale=.60]{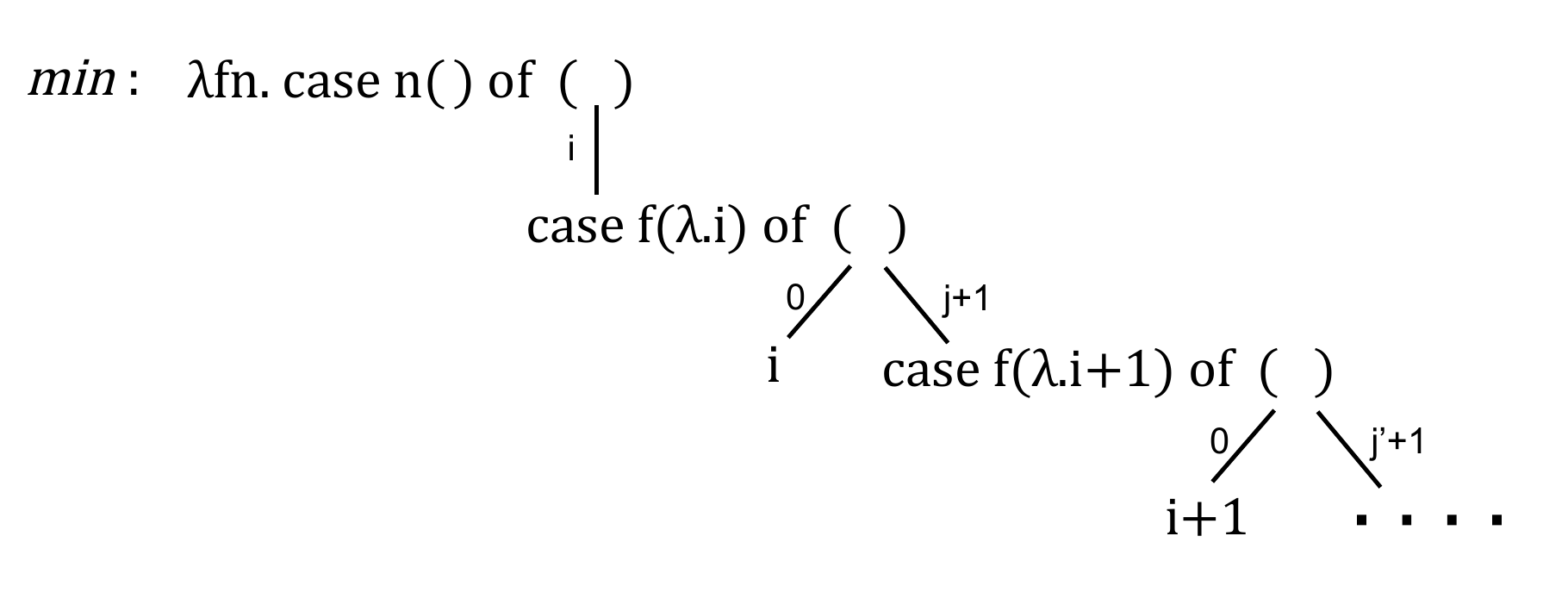}
\vspace*{-1.5ex}
\includegraphics[scale=.60]{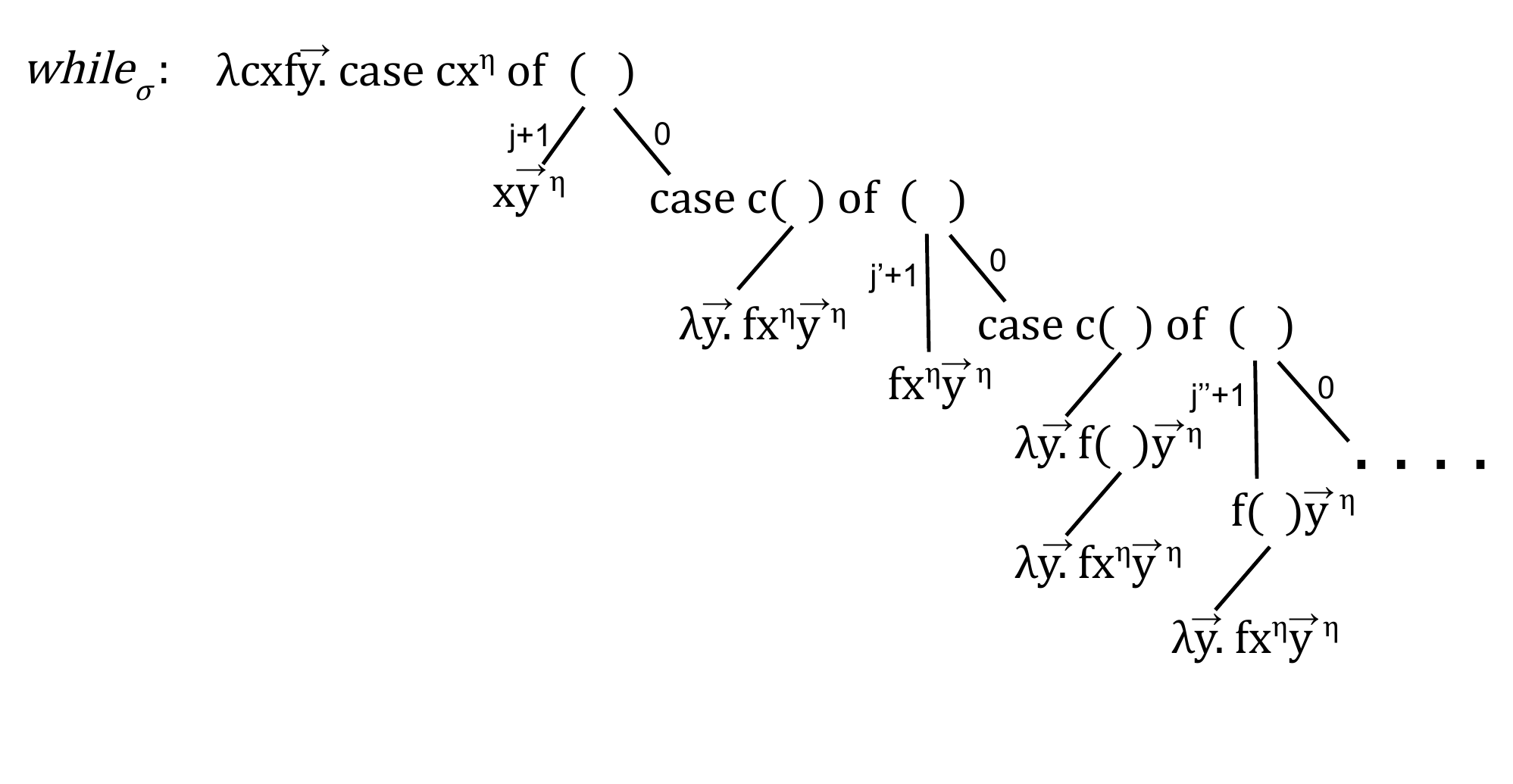}
\end{center}

\vspace*{-5.0ex}
Clearly, each of these NSPs is left-well-founded; the one for $\mmin$ is even left-bounded,
since there is no nesting of calls to $n$ or $f$.
Since also the NSPs for 
$\suc$, $\pre$, $\ifzero$, $\byval$ and $x^{\sigma\eta}$ are plainly LWF, 
and LWF procedures are closed under $\lambda$-abstraction and application 
(Theorem~\ref{substructures-thm}(i)), we have the following important result:

\begin{theorem}  \label{Tmin-LWF-thm}
The interpretation of any term of $\sysT+\mmin$ or $\sysW$ is an LWF procedure.
\end{theorem}

The reader may also enjoy constructing the appropriate trees for
$\rec_\sigma^\str$ and $\while_\sigma^\str$ where $\sigma$ is of level $0$,
and to observe that these trees are \emph{left-bounded} 
(in contrast to those for $\rec_\sigma$ and $\while_\sigma$).
Since the trees for $\mmin$ and all the constants of $B$ are also left-bounded,
we may infer by Theorem~\ref{substructures-thm}(ii)
that the interpretation of any term of $\sysT_0^\str + \mmin$ or $\sysW_0^\str$ is left-bounded.
This may shed light on the discussion of Section~\ref{sec-intro}, 
but will not be formally required for the remainder of the paper.

\subsection{Bar recursors}  \label{sec-bar-recursors}

We conclude the section by explaining the notions of bar recursion that we shall use.
Traditionally, bar recursion has usually been considered either as a purely syntactic operation 
(as in \cite{Spector-bar}), or as an element of a total type structure such as the 
Kleene-Kreisel \emph{total continuous} functionals $\Ct$
or Bezem's \emph{strongly majorizable} functionals $\mathsf{SM}$.
Here, for expository purposes, we shall introduce bar recursion 
first in a `naive' way with reference to the full set-theoretic type structure $\Set$,
and then as an operation within $\SP^0$ defined by a certain $\PCF$ program---%
the latter will provide the setup for the main theorem of Section~\ref{sec-main-theorem}.
In Section~\ref{sec-other-models}, we will relate this to more familiar notions of bar recursion
and will show how our results transfer relatively easily from $\SP^0$ to models 
such as $\Ct$ and $\mathsf{SM}$.

As explained in Section~\ref{sec-intro}, bar recursion is in essence recursion over well-founded trees. 
For the purpose of this paper, a \emph{tree} $\TT$ will be an inhabited prefix-closed subset of $\N^*$
(the set of finite sequences over $\N$), 
with the property that for every $\vec{x} = (x_0,\ldots,x_{i-1}) \in \TT$, one of the following holds:
\begin{enumerate}
\item There is no $y \in \N$ such that $(x_0,\ldots,x_{i-1},y) \in \TT$ 
(we then say $\vec{x}$ is a \emph{leaf} of $\TT$).
\item For all $y \in \N$, $(x_0,\ldots,x_{i-1},y) \in \TT$
(we then say $\vec{x}$ is an \emph{internal node} of $\TT$).
\end{enumerate}
We write $\TT^l, \TT^n$ for the set of leaves and internal nodes of $\TT$ respectively.

A tree $\TT$ is \emph{well-founded} if there is no infinite sequence $x_0,x_1,\ldots$
over $\N$ such that $(x_0,\ldots,x_{i-1}) \in \TT$ for every $i$. 
Thus, in a well-founded tree, every maximal path terminates in a leaf.
If $\TT$ is well-founded, a function $f: \TT \arrow \N$ may be defined by recursion on the tree structure 
if we are given the following data:
\begin{itemize}
\item A \emph{leaf function} $L: \TT^l \arrow \N$ specifying the value of $f$ on leaf nodes.
\item A \emph{branch function} $G: \TT^n \times \N^\N \arrow \N$ specifying the value of $f$ on
an internal node $\vec{x}$, assuming we have already defined the value of $f$ on all the immediate 
children of $f$.
Specifically, if $b(y)$ gives the value of $f(\vec{x},y)$ for every $y \in \N$, then 
$G(\vec{x},b)$ gives the value of $f(\vec{x})$.
\end{itemize}
Indeed, we may define the function $BR^\TT_{L,G}$ obtained by \emph{bar recursion} from $L$ and $G$
to be the unique function $\TT \arrow \N$ satisfying:
\begin{eqnarray*}
BR^\TT_{L,G}(\vec{x}) & = & L(\vec{x})  \mbox{~~~~if $\vec{x} \in \TT^l$} \\
BR^\TT_{L,G}(\vec{x}) & = & G(\vec{x},\;\Lambda y.\, BR^\TT_{L,G}(\vec{x},y))  \mbox{~~~~if $\vec{x} \in \TT^n$}
\end{eqnarray*}
The existence and uniqueness of $BR^\TT_{L,G}$ are easy consequences of well-foundedness.

It is easy to see how $L$ and $G$ may be represented by objects of simple type:
elements of $\TT$ may be represented by elements of $\N$ via some standard primitive recursive coding
$\ang{\cdots}: \N^* \arrow \N$, so that we may consider $L$ and $G$ as functions
$\N \arrow \N$ and $\N \times \N^\N \arrow \N$ respectively.
As regards the tree $\TT$ itself, we now introduce two related ways, due respectively to
Spector \cite{Spector-bar} and Kohlenbach \cite{Kohlenbach-bar-rec}, for representing
well-founded trees by means of certain functionals $F: \N^\N \arrow \N$.
We shall write $|\vec{x}|$ for the length of a sequence $\vec{x}$,
and if $j \in \N$, shall write $[\vec{x}\, j^\omega]$
for the primitive recursive function $\N \arrow \N$ defined by
\[ [x_0,\ldots,x_{r-1},j^\omega](i) ~=~ \left\{ \begin{array}{ll}
    x_i & \mbox{if $i<r$,} \\
    j  & \mbox{if $i \geq r$} \end{array} \right.\]

\begin{definition}  \label{leaf-def}
Suppose $F$ is any function $\N^\N \arrow \N$.

(i) We say $\vec{x} \in \N^*$ satisfies the \emph{Spector bar condition} (with respect to $F$) if 
\[ F([\vec{x}\, 0^\omega]) ~<~ |\vec{x}| \;, \]
and the \emph{Kohlenbach bar condition} if
\[ F([\vec{x}\, 0^\omega]) ~=~ F([\vec{x}\, 1^\omega]) \;. \]

(ii) The \emph{Spector tree} of $F$, written $\TT^S(F)$, is the set of sequences $\vec{x} \in \N^*$
such that no proper prefix of $\vec{x}$ satisfies the Spector bar condition w.r.t.\ $F$. 
The \emph{Kohlenbach tree} $\TT^K(F)$ is defined analogously using the Kohlenbach bar condition.
\end{definition}

Both $\TT^S(F)$ and $\TT^K(F)$ are clearly trees in our sense. 
Furthermore, the following important fact ensures a plentiful supply of functionals giving rise to
\emph{well-founded} trees.

\begin{proposition}  \label{continuous-wf-prop}
If $F$ is continuous with respect to the usual Baire topology on $\N^\N$, 
then both $\TT^S(F)$ and $\TT^K(F)$ are well-founded.
\end{proposition}

\noindent \proofsk
For any infinite sequence $x_0,x_1,\ldots$, there will be some `modulus of continuity' $m$ for $F$ such that 
for all $n \geq m$ and all $j$ we have $F([x_0,\ldots,x_{n-1},j^\omega]) = F(\Lambda i.x_i)$.
It follows easily that some finite subsequence $(x_0,\ldots,x_{n-1})$ will satisfy 
the Spector [resp.\ Kohlenbach] condition. 
The shortest such prefix will then be a leaf in $\TT^S(F)$ [resp.\ $\TT^K(F)$].
\QED

\vspace*{1.5ex}
There are also other ways in which well-founded trees may arise---for instance, it is well-known that
if $F$ is \emph{majorizable} then $\TT^S(F)$ is well-founded---but it is the continuous case that will
be most relevant to our purposes.

Using the above representations, we may now introduce our basic definition of bar recursion.
We are here naively supposing that $F,L,G$ are drawn from the full set-theoretic type structure $\Set$,
although this is not really the typical situation;
relativizations of this definition to other total type structures 
will be considered in Section~\ref{sec-other-models}.

\begin{definition}[Bar recursors]  \label{bar-rec-def}
A \emph{Spector bar recursor} (over $\Set$) is any partial function 
\[ BR ~:~ \Set((\nat\arrow\nat)\arrow\nat) \;\times\; \Set(\nat\arrow\nat) \;\times\; \Set(\nat \times (\nat\arrow\nat) \arrow \nat) ~\parrow~ \N  \]
such that for all $F : \N^\N \arrow \N$ with $\TT^S(F)$ well-founded
and for any $L : \N \arrow \N$ and $G: \N \times \N^\N \arrow \N$, we have
\begin{eqnarray*}
BR(F,L,G)(\ang{\vec{x}}) & = & L(\ang{\vec{x}}) \mbox{~~~whenever $\vec{x} \in \TT^S(F)^l$,} \\
BR(F,L,G)(\ang{\vec{x}}) & = & G(\ang{\vec{x}},\;\Lambda z.\,BR(F,L,G)(\ang{\vec{x},z})
    \mbox{~~~whenever $\vec{x} \in \TT^S(F)^n$}.
\end{eqnarray*}
The notion of Kohlenbach bar recursor is defined analogously using $\TT^K(F)$.
\end{definition}
Note that the above equations uniquely fix the value of $BR(F,L,G)(\ang{\vec{x}})$ for all 
$\vec{x} \in \TT^S(F)$.
We shall not be concerned with the behaviour of bar recursors $BR$ on sequences $\vec{x}$
outside the tree in question,%
\footnote{In this respect our definition of `bar recursor' here is slightly weaker than some given in the literature
(e.g.\ in \cite[Section~7.3]{HOC}); this will in principle make our main theorem slightly stronger, though
not in any deep or essential way.}
nor with their behaviour on arguments $F$ such that $\TT(F)$ is not well-founded.

Our next step will be to re-construe the definition of bar recursors as a recursive program within $\PCF$.
For this, we first note that both the Spector and Kohlenbach bar conditions are readily testable 
by means of programs in $\PCF$ or indeed in $\sysT_0^\str$.
Although these languages have just a single base type $\nat$, for clarity we shall write $\nat^*$ for
occurrences of $\nat$ whose role is to represent sequences $\vec{x}$ via their codes $\ang{\vec{x}}$.
We shall suppose we have a fixed choice of $\sysT_0^\str$ programs
\[ \len ~:~ \N^* \arrow \N \;, ~~~~~~~~
   \append ~:~ \N^* \arrow \N \arrow \N^* \;, ~~~~~~~~
   \basic ~:~ \N^* \arrow \N \arrow (\N \arrow \N)  \]
such that for any $\vec{x},z,j,i$ we have
\[ \len\;\ang{\vec{x}} ~\reducesto^*~ |\vec{x}| \;, ~~~~~~~~
   \append\;\ang{\vec{x}}\;z ~\reducesto^*~ \ang{\vec{x},z} \;, ~~~~~~~~
   \basic\;\vec{x}\;j\;i ~\reducesto^*~ [\vec{x},j](i) \;,
\]
(where we omit the hats from PCF numerals).
We also presuppose fixed $\sysT_0^\str$ implementations of $=$ and $<$.
Using this machinery, we may now define a $\PCF$ term
\[ \BR^S ~:~ ((\nat \arrow \nat) \arrow \nat) ~\arrow~ 
                        (\nat^* \arrow \nat) ~\arrow~
                        (\nat^* \arrow (\nat \arrow \nat) \arrow \nat) ~\arrow~
                        (\nat^* \arrow \nat) \]
by 
\begin{eqnarray*}
\BR^S\;F\;L\;G\;x & = & \iif~(F (\basic(x,0)) < \len\;x)~\tthen~L\;x \\
                           &     & \eelse~G\;x\;(\lambda z.\,\BR^S\;F\;L\;G\;(\append\;x\;z)) 
\end{eqnarray*}
or a little more formally by
\begin{eqnarray*}
\BR^S & = & \lambda FLG.\;Y_{\nat\arrow\nat}\, (\lambda B.\;\lambda x.\, \\
          &     & ~~~~~~\iif~(F (\basic(x,0)) < \len\;x)~\tthen~L\;x~\eelse~G\;x\;(\lambda z.\,B (\append\;x\;z))) \;.
\end{eqnarray*}
The Kohlenbach version $\BR^K$ (of the same type) is defined analogously, replacing the subterm
$(F (\basic(x,0)) < \len\;x)$ by $(F (\basic(x,0)) = F (\basic(x,1)))$.

Both of these PCF terms may be interpreted in $\SP^0$, yielding NSPs at the above type which we shall
also denote by $\BR^S$ and $\BR^K$ respectively.
We shall refer to these as the \emph{canonical (Spector or Kohlenbach) bar recursors} within $\SP^0$.

It is not hard to see intuitively that these NSPs for $\BR^S$ and $\BR^K$ are non-LWF, since the
unrolling of the recursion will lead to an infinite sequence of nested calls to $G$.
To illustrate the phenomenon, we schematically depict here part of the NSP for 
$\lambda FLG.\;\BR^S\,F\,L\,G\,\ang{\,}$ (so that evaluations of $x$ are elided):

\vspace*{-0.5ex}
\begin{center}
\includegraphics[scale=.60]{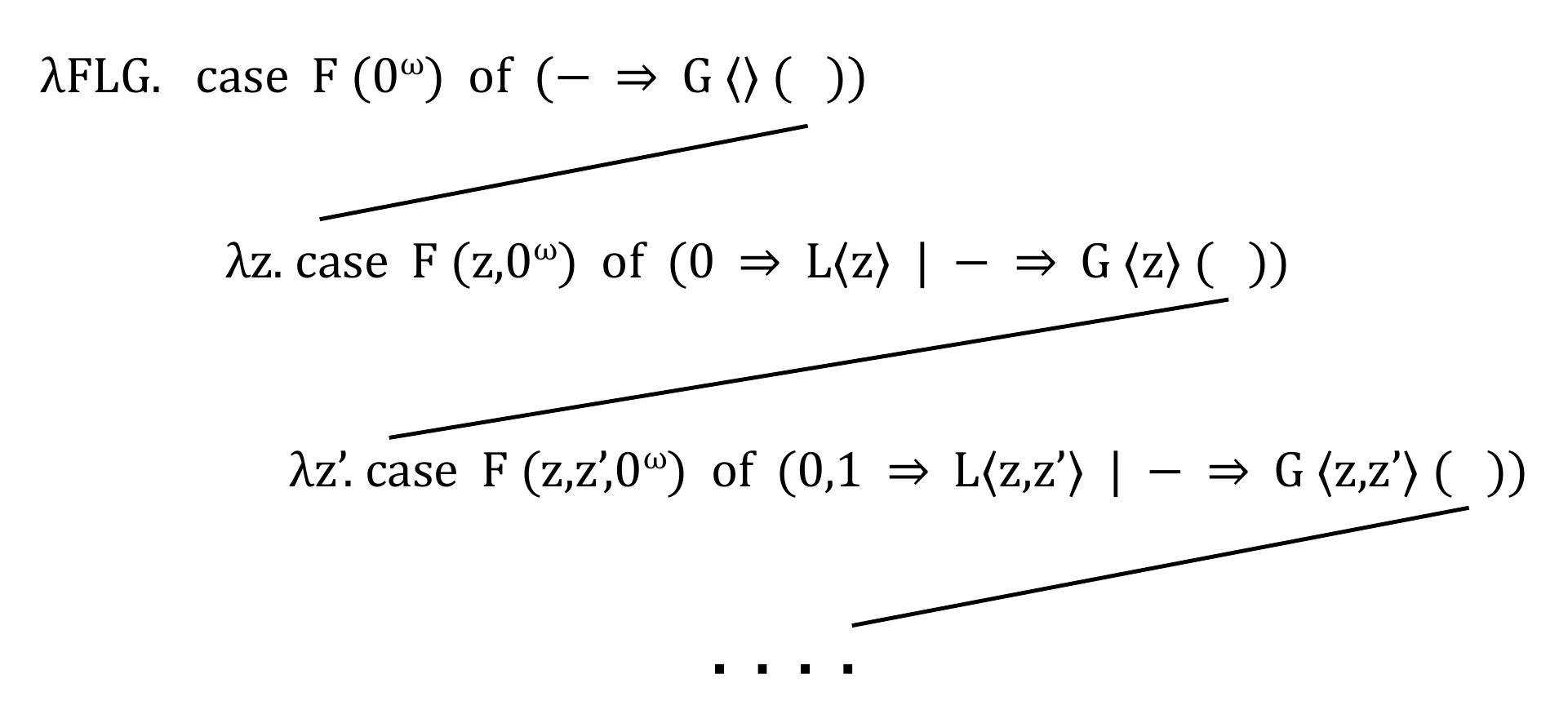}
\end{center}

It follows immediately by Theorem~\ref{Tmin-LWF-thm} that 
these particular bar recursors within $\SP^0$ are not denotable in $\sysT+\mmin$.
However, this does not in itself address the main question of interest:
what we wish to know is that no element of $\SP^0$ can have the extensional behaviour of a bar recursor,
even if we restrict attention to arguments $F,L,G$ which represent \emph{total} functionals in the spirit of
Definition~\ref{bar-rec-def}.
This begs the question of what it means for an element of $\SP^0$ to `represent' a total functional.
We now briefly indicate why there is room for several reasonable answers to this question, 
thus motivating our `robust' approach which is designed to work for all of them.

The general picture we have in mind is that of some chosen type structure $T$ 
of total functionals over $\N$---%
that is, a family of sets $T(\sigma)$ where $T(\nat) = \N$ and $T(\sigma\arrow\tau)$ is some set
of total functions $T(\sigma) \arrow T(\tau)$---along with some way of representing $T$ within $\SP^0$.
The latter will in general consist of what in \cite{HOC} we call a (type- and numeral-respecting) 
\emph{applicative simulation} $\gamma: T \rlzn \SP^0$: that is, a family of total relations
$\gamma_\sigma \subseteq T(\sigma) \times \SP^0(\sigma)$ such that $\gamma_N(n,x)$ iff $x=n$,
and $\gamma_{\sigma\arrow\tau}(f,f')$ and $\gamma_\sigma(x,x')$ imply $\gamma(f(x), f' \cdot x')$.
Relative to this, we may call an element of $\SP^0(\sigma)$ \emph{total} 
if it is in the image of $\gamma_\sigma$.
We might then regard as a `bar recursor' in $\SP^0$ any procedure $\Phi$ that satisfies (an analogue of)
the equations of Definition~\ref{bar-rec-def} for all total $F,L,G \in \SP^0$ of appropriate types.

However, there is scope for variation here, both in the choice of $T$ (which could be either
$\Ct$ or $\HEO$, for example) and in the choice of the simulation $\gamma$.
Different choices will in general lead to different notions of `total element' within $\SP^0$
(this phenomenon is explored in \cite{Plotkin-totality}), and hence to different criteria for what it means
to be a bar recursor in $\SP^0$.

Our approach to dealing with this is to identify a core class of elements of $\SP^0$ which are likely
to be `total' under all reasonable choices of interest, and then postulate, as a minimal requirement
for any proposed `bar recursor' in $\SP^0$, that the equations of Definition~\ref{bar-rec-def}
should at least be satisfied by these core total elements. 
Our main theorem will then claim that even this minimal requirement cannot be met by any LWF procedure.
As we shall argue in Section~\ref{sec-other-models}, this will enable us to conclude that in no reasonable
sense can a bar recursor be definable in $\sysT+\mmin$ or $\sysW$.

In fact, a suitable class of core total elements for our purpose will be those definable in
the language $\sysT_0^\str$ of Subsection~\ref{sec-weaker-langs}
(interpreted in $\SP^0$ via its translation to $\PCF$).  
Our rationale for this is that all reasonable choices of the total type structure $T$ can be expected to
be models for $\sysT_0^\str$ (or equivalently for Kleene's S1--S8), 
and that it is furthermore a mild requirement
that $\gamma$ should relate the interpretation of any closed $\sysT_0^\str$ term in $T$ to its interpretation in
$\SP^0$. (This will in fact be so if it is the case for the standard programs $k$ and $s$ and for the 
constants $\suc$, $\pre$, $\ifzero$, $\rec_\sigma^\str$.)
Our thesis, then, is that the $\sysT_0^\str$ definable elements of $\SP^0$ can be expected to be `total'
in all senses of interest.

All of this leads us to the following definitions.
Here we identify the PCF programs $\basic$, $\append$, $\len$ with their interpretations in $\SP^0$.

\begin{definition}  \label{NSP-SK-trees}
Suppose $F \in \SP^0((\nat\arrow\nat)\arrow\nat)$

(i) We say $\vec{x} \in \N^*$ satisfies the Spector bar condition with respect to $F$ if 
$F \cdot (\basic \cdot \ang{\vec{x}} \cdot 0) < \len \cdot \ang{\vec{x}}$,
and the Kohlenbach bar condition if
$F \cdot (\basic \cdot \ang{\vec{x}} \cdot 0) = F \cdot (\basic \cdot \ang{\vec{x}} \cdot 1)$.

(ii) The tree $\TT^S(F)$ [resp.\ $\TT^K(F)$] consists of all sequences $\vec{x}$ such that no proper
prefix of $\vec{x}$ satisfies the Spector [resp.\ Kohlenbach] bar condition.
\end{definition}

The following easy fact will be useful:

\begin{proposition}
If $F \in \SP^0((\nat\arrow\nat)\arrow\nat)$ is definable by a term of $\sysT_0^\str$,
then $\TT^S(F)$, $\TT^K(F)$ are well-founded.
\end{proposition}

\noindent \proof
If $F$ is $\sysT_0^\str$ definable, clearly $F$ will represent a total functional 
$\overline{F}: \N^\N \arrow \N$ with respect to the obvious representation of functions $\N \arrow \N$ 
within $\SP^0(\nat\arrow\nat)$.
Moreover, since application in $\SP^0$ is continuous, it is easy to see that $\overline{F}$ will be
continuous for the Baire topology, and so by Proposition~\ref{continuous-wf-prop}
the trees $\TT^S(F) = \TT^S(\overline{F})$ and $\TT^K(F) = \TT^K(\overline{F})$ will be well-founded.
\QED

\begin{definition}  \label{weak-bar-rec-def}
A \emph{weak Spector bar recursor} in $\SP^0$
is an element 
\[ \Phi ~\in~ \SP^0~ (((\nat \arrow \nat) \arrow \nat) ~\arrow~ 
               (\nat^* \arrow \nat) ~\arrow~
               (\nat^* \arrow (\nat \arrow \nat) \arrow \nat) ~\arrow~
               (\nat^* \arrow \nat)) \]
such that the following hold for all $\sysT_0^\str$-definable $F,L,G \in \SP^0$
of appropriate types such that for all $\vec{x} \in \TT^S(F)$:
\[ \begin{array}{rcl}
\Phi \cdot F \cdot L \cdot G \cdot \ang{\vec{x}} & = & 
   L \cdot \ang{\vec{x}}
   \mbox{~~~if $\vec{x} \in \TT^S(F)^l$} \;, \\
\Phi \cdot F \cdot L \cdot G \cdot \ang{\vec{x}} & = & 
   G \cdot \ang{\vec{x}} \cdot 
   (\lambda z^\nat.\; \Phi \cdot F \cdot L \cdot G \cdot (\append \cdot \ang{\vec{x}} \cdot z)) 
   \mbox{~~~if $\vec{x} \in \TT^S(F)^n$} \;.
\end{array} \]
The notion of \emph{weak Kohlenbach bar recursor} in $\SP^0$ is defined analogously.
\end{definition}
Here the abstraction $\lambda z$ is understood simply as a $\lambda$-abstraction within the language
of NSPs; note that for any given $\Phi,F,L,G,\vec{x}$, the body of this abstraction will evaluate to an 
NSP with free variable $z$.

Clearly the canonical bar recursors $\BR^S$, $\BR^K$ defined earlier are examples of weak bar recursors
in this sense. Moreover:

\begin{proposition}  \label{weak-br-defined-prop}
If $\Phi$ is any weak (Spector or Kohlenbach) bar recursor, $F,L,G$ are $\sysT_0^\str$-definable,
and $\vec{x} \in \TT(F)$,
then the value of $\Phi \cdot F \cdot L \cdot G \cdot \ang{\vec{x}}$ is a numeral 
and is uniquely determined by the defining equations above.
\end{proposition}

\noindent \proof
For a given $F,L,G$, we show that the set $S$ of $\vec{x} \in \TT(F)$ for which the proposition holds 
contains all leaves and all internal nodes whose immediate children are all in $S$;
it follows that $S$ is the whole of $\TT(F)$ since the latter is well-founded.
For the step case, we use the fact that if $G$ is $\sysT_0^\str$-definable and 
$h \in \SP^0(\nat\arrow\nat)$ is total (i.e.\ $h \cdot z \in \N$ for all $z \in \N$),
then $G \cdot x \cdot h \in \N$ for any $x \in \N$.
We show this by an easy induction on the $\sysT_0^\str$ term that denotes $G$,
which we may assume to be some $\beta$-normal form $\lambda xh.M$,
so that all variables are bound within $M$ are of type level 0.
\QED

\vspace*{1.5ex}
Clearly, if $F$ represents an element $\check{F} \in \Ct((\nat\arrow\nat)\arrow\nat)$ 
via any reasonable simulation $\gamma$, it will be automatic that $\TT(F)$ is well-founded since any such
$\check{F}$ is continuous. Indeed, since bar recursors exist as third-order functionals $\check{\Phi}$ 
within $\Ct$, any elements $\Phi \in \SP^0$ that represent such $\check{\Phi}$ will be \emph{total}
weak bar recursors relative to $\gamma$.
The situation is different for the type structure $\HEO$:
there are classically discontinuous functions $\check{F} \in \HEO((\nat\arrow\nat)\arrow\nat)$,
and if $F \in \SP^0$ represents such an $\check{F}$ then $\TT(F)$ may be non-well-founded,
in which case Definition~\ref{weak-bar-rec-def} places no condition on how $\Phi$ should behave on $F$.

In our main proof, we shall find it more convenient to work with the Kohlenbach definition, 
but the theorem will transfer readily to the Spector version 
in view of the following easy relative definability result.
From here on, we shall allow ourselves to write $k$ for the pure type of level $k$, so that
$0$ denotes $\nat$ and $k+1$ denotes $k \arrow \nat$.

\begin{proposition}  \label{spect-kohlen-prop}
If $\Phi^S$ is any weak Spector bar recursor in $\SP^0$, 
then a weak Kohlenbach bar recursor $\Phi^K$ is $\sysT+\mmin$ definable 
relative to $\Phi^S$.
Hence if an LWF weak Spector bar recursor exists in $\SP^0$, 
then so does an LWF weak Kohlenbach bar recursor.
\end{proposition}

\noindent \proof~
We first construct a $\sysT+\mmin$ definable element
$U \in \SP^0(2 \arrow 2)$ such that for any $F \in \SP^0(2)$ 
whose Kohlenbach tree is well-founded and is not simply $\{ \ang{\,} \}$,
the Spector tree of $U \cdot F$ is precisely the Kohlenbach tree of $F$.
We may achieve this by defining (in PCF-style notation)
\[ U ~=~ \lambda F.\lambda g.\;(\mmin\;r.\; F([g(0),\ldots,g(r-1),0^\omega]) =
       F([g(0),\ldots,g(r-1),1^\omega])) - 1 \;. \]
Using this, we may define
\[ \Phi^K ~=~ \lambda FLGx.~\iif\;F([0^\omega])=F([1^\omega])\;\tthen\;L\ang{}
             \;\eelse\;\Phi^S(U(F),L,G)(x) \;. \]
It is now easy to check by bar induction on nodes in $\TT^K(F)$
that $\BR^K$ is a weak Kohlenbach bar recursor.
\QED

\vspace*{1.0ex}
Conversely, a more subtle argument (given in Kohlenbach \cite{Kohlenbach-bar-rec}) 
shows that Spector bar recursion is definable from Kohlenbach bar recursion even in System~$\sysT$,
though we shall not need this here.
We also refer the reader to Escard\'o and Oliva \cite{Bar-rec-prod-selection} for a cornucopia 
of related functionals known to be either interdefinable with or stronger than Spector bar recursion
over System~$\sysT$; our main theorem will thus yield that none of these functionals are
definable in $\sysT + \mmin$.

One final preliminary is needed.
In order to ease notation in our main proof, we shall actually consider a simpler kind of bar recursor
readily obtained as a specialization of those described above.

\begin{definition}  \label{restricted-def}
A \emph{simplified weak Spector bar recursor} (in $\SP^0$) is an element
\[ \Phi \in \SP^0(2 \arrow 2 \arrow 1) \]
such that the following hold for all $T_0^\str$-definable $F,G \in \SP^0$
of appropriate types such that for all ${\vec{x}} \in \TT^S(F)$:
\[ \begin{array}{rcl}
\Phi \cdot F \cdot G \cdot \ang{\vec{x}} & = & 
   2 \ang{\vec{x}}+1 \mbox{~~~if $\vec{x} \in \TT^S(F)^l$} \;, \\
\Phi \cdot F \cdot G \cdot \ang{\vec{x}} & = & 
       G \cdot (\lambda z^\nat.\; \Phi \cdot F \cdot G \cdot (\append \cdot \vec{x}\ \cdot z))
        \mbox{~~~if $\vec{x} \in \TT^S(F)^n$} \;.
\end{array} \]
The notion of \emph{simplified weak Kohlenbach bar recursor} is defined analogously.
\end{definition}

It is easily seen that a simplified weak (Spector or Kohlenbach) bar recursor
is $\sysT_0^\str$-definable from an ordinary one 
just by specializing the leaf function $L$ to $\lambda x.2x+1$
(this move is admittedly hard to motivate at this point!)
and by eschewing the dependence of $G$ on an argument $x$.
By analogy with Proposition~\ref{weak-br-defined-prop}, we have:

\begin{proposition}  \label{simplified-weak-br-defined-prop}
If $\Phi$ is any simplified weak (Spector or Kohlenbach) bar recursor, $F,G$ are $\sysT_0^\str$-definable,
$\TT(F)$ is well-founded and $\vec{x} \in \TT(F)$,
then the value of $\Phi \cdot F \cdot G \cdot \ang{\vec{x}}$ is a numeral 
and is uniquely determined by the defining equations above.
\end{proposition}

The proof of Proposition~\ref{spect-kohlen-prop} clearly also yields the following:

\begin{proposition}  \label{restricted-spect-kohlen-prop}
If an LWF simplified weak Spector bar recursor exists in $\SP^0$, 
so does an LWF simplified weak Kohlenbach bar recursor.
\end{proposition}

Such simplified bar recursors were called \emph{restricted} bar recursors in \cite[Section~6.3.4]{HOC},
but the latter name clashes with a different use of the same term by Spector in \cite{Spector-bar}.
In $\cite{HOC}$ we supposed that the simplified bar recursors were weaker than the general ones,
so that they led to a slightly stronger non-definability result.
Actually, it turns out to be not too hard to define general bar recursors from simplified ones
(we leave this as an exercise for the interested reader).
Nevertheless, we shall prove our main theorem for the simplified versions,
both because that was what was claimed in \cite{HOC}, and because it does lighten the notational load
in parts of our proof.
Against this, the later parts of the proof (Sections~\ref{sec-ctex} and \ref{sec-properties}) 
turn out to be a little more delicate in the simplified setting, but we think there is also some interest
in the opportunity this gives for illustrating the versatility of our method of proof.

\section{The main theorem}  \label{sec-main-theorem}

In this section, we shall prove the following theorem: 

\begin{theorem}  \label{main-theorem}
\emph{Within $\SP^0$, no simplified weak Kohlenbach bar recursor can be LWF,
and hence none can be definable in $\sysT + \mmin$ or in $W$.}
\end{theorem}

The corresponding fact for simplified Spector bar recursion 
(stated as Theorem~6.3.28 in \cite{HOC}) 
will then follow immediately by Proposition~\ref{restricted-spect-kohlen-prop}.
It will also follow, \textit{a fortiori}, that no ordinary Spector or Kohlenbach bar recursor
in the sense of Definition~\ref{bar-rec-def} can be LWF.
From here on we shall consider only Kohlenbach bar recursion, 
and will write $\TT^K(F)$ simply as $\TT(F)$.

The proof of Theorem~\ref{main-theorem} follows the method of proof
of Theorem~6.3.27 in \cite{HOC}, which shows that 
no NSP weakly representing the System~T recursor $\rec_{\nat\arrow\nat}$ 
can be definable in $\sysT_0^\str + \mmin$.
The proof of this theorem is already quite intricate, and 
that of the present theorem adds some further ingredients. 
The reader may therefore find it helpful to study the proof of
\cite[Theorem~6.3.27]{HOC} in conjunction with the present one---%
however, the account given here will be technically self-contained,
and we shall also offer a more extended motivational discussion here than we did in \cite{HOC}.%
\footnote{We take the opportunity to draw attention here to a small error in the proof of
\cite[Theorem~6.3.27]{HOC}. 
In the penultimate sentence of the proof (on page 258), we claim that $F_1(g_{()}^{F_1}) = K$.
For this, we use the previously established fact that $g_{()}^{F_1}(0) = \ang{0,\ldots,0}$;
in the context of the proof, it is natural to denote this by $y_0$. However, one also wants that $y_0$
is distinct from all of $y_1,\ldots,y_d$, which we do not here know to be the case.

The problem is readily fixed by simply adding, at the point at which each $y_w$ is selected 
(for $1 \leq w \leq d$), the further requirement that $y_w \neq y_0$. For this, we need to add 1 to the
lower bounds on the moduli $m^w$ from earlier in the proof, so that we take $m^0 > n^0 + 2$,
$m^1 > n^0 + n^1 + 3$, etc.
Then, when picking the path through the tree for $\Psi_d$ at the bottom of page 256, we should start by
defining $y_0 = g_{()}^{F_0}(0)$ (so that actually $y_0 = c$), then insert the requirements that 
$y_1,y_2,\ldots$ differ from $y_0$. (In the last line of page 256, $y_0$ was originally intended to read $y_1$,
but should now be modified to $y_0,y_1$.)

In the third-to-last line of page 257, the claim that $g^{F_1}_{(\vec{z},0)}(0) = y_w$ now holds
even when $w=0$. }

We start with an informal outline of our argument.
Suppose that $\Phi$ is any genuine (simplified, weak) bar recursor
as per Definition~\ref{restricted-def}, and that $\Psi$ is some LWF procedure
purported to be such a bar recursor.
We shall set $\Phi_0 = \lambda FG.\, \Phi F G \ang{}$ and $\Psi_0 = \lambda FG.\, \Psi F G \ang{}$,
so that $\Psi_0$ is also LWF by  Theorem~\ref{substructures-thm}(i).
Our task will be to find some particular $\sysT_0^\str$-definable arguments 
$F \in \SP^0(2)$, $G \in \SP^0(2)$ such that $\Psi_0 \cdot F \cdot G \neq \Phi_0 \cdot F \cdot G$.
Since $\Phi$ here is an arbitrary bar recursor, this will show that $\Psi$ is not a bar recursor after all.


The key idea is that $\Psi$, being LWF, will only be prepared to nest calls to $G$ to a finite depth
along any specified computation path.
On the other hand, $\Phi$, being a genuine bar recursor, must be willing to nest such calls to any
depth required, as dictated by $\TT(F)$.
In order to manifest an extensional difference between $\Phi_0$ and $\Psi_0$, 
we therefore wish to construct an $F$ such that $\TT(F)$
that goes deeper on some path than $\Psi_0$ is willing to explore, 
together with a $G$ that forces the computation of $\Phi_0$ to explore precisely this path.
In this way, we can arrange that the computation of $\Phi_0 \cdot F \cdot G$ retrieves from
within $G$ some numerical value $K$ that is not discoverable by $\Psi_0$, and propagates it to the top level.

Much of the proof is aimed at acquiring a sufficient grasp of the behaviour of $\Psi_0$ that we can
guarantee that $\Psi_0 \cdot F \cdot G$ does \emph{not} return this value $K$.
Our approach to this will be similar to that in \cite[Theorem~6.3.27]{HOC}.
Assuming for the moment that we know how to obtain a suitable $F$,
we shall start by considering the computation of $\Psi_0 \cdot F \cdot G_0$
for a certain very simple functional $G_0$. Suppose that this evaluates to some number $c$.
By analysing this and some related computations in detail, we shall discover a set of properties of $G_0$
that suffice to \emph{secure} this computation, in the sense that for any other $G$ with these 
properties, a precisely similar computation will go through, yielding the same result $c$.
Put another way, we shall find a certain \emph{neighbourhood} $\Gcal$ of $G_0$ such that
for all $G \in \Gcal$ we have $\Psi_0 \cdot F \cdot G = c$.
Moreover, the construction of $\Gcal$ will be so arranged that
it is possible to pick some $G_1 \in \Gcal$ which forces $\Phi_0$ to explore beyond 
the reach of $\Psi_0$ in the manner suggested above.
Indeed, by choosing such a $G_1$ with some care, we can ensure that
$\Phi_0 \cdot F \cdot G_1$ evaluates to a number $K$ chosen to be different from $c$.
This establishes the required difference between $\Phi_0$ and $\Psi_0$.

The main new ingredient, not present in the proof of \cite[Theorem~6.3.27]{HOC},
concerns the way in which a suitable argument $F$ is chosen.
As indicated above, we want $F$ to represent a well-founded tree that `undercuts' the tree explored by
$\Psi_0$ in a certain computation; on the other hand, the computation performed by $\Psi_0$ will
itself depend partly on the argument $F$ that we give it.
This apparent circularity suggests that we should try to arrive at a suitable $F$ (which we call $F_\infty$)
by a process of successive approximation in tandem with our analysis 
of the computation of $\Psi_0 \cdot F \cdot G_0$.
This will allow us to ensure that $\TT(F_\infty)$ undercuts $\Psi_0$ with respect to the computation of
$\Psi_0 \cdot F_\infty \cdot G_0$ itself.

More specifically, our proof will be structured as follows.
In Section~\ref{3.1} we begin with some very simple functionals $G_0 \in \SP^0(2)$ 
and $F_0^+ \in \SP^0(2)$, 
of which the latter will serve as the first step in the iterative construction of a suitable $F$.
By analysing the computation of $\Psi_0 \cdot F_0^+ \cdot G_0 = c$, initially just at the `top level'
(that is, without delving into the computations of the type~1 arguments passed to $F_0^+$ and $G_0$), 
we are able to glean some `neighbourhood information' about $G_0$ which helps to secure aspects of 
this computation (and will also secure the corresponding computation for $G_1$ once the latter has
been constructed). In the course of this, we will also have replaced $F_0^+$ by the next iteration $F_1^+$.

However, the information about $G_0$ gathered so far does not by itself suffice to secure the entire
computation: the top-level computation will typically rely on certain information about 
the arguments passed to $F$ and $G$; and since these may themselves involve calls to $G$,
some further constraints on $G_0$ may be needed to secure this information.
We are thus led to repeat our analysis for certain subcomputations associated with the
arguments to $F$ and $G$---and so on recursively to whatever depth is required.
This is done in Section~\ref{3.2}.

A key step in the proof is to observe that since $\Psi_0$ is LWF, this entire construction will eventually
bottom out in a situation where no further subcomputations need to be analysed (there is a crucial
appeal to K\"onig's lemma here). We record what happens at this final stage of the construction
in Section~\ref{3.3}.

At the end of this computation analysis, we are left with two things.
First, in the course of the analysis, the value of $F$ we are considering
will have been successively refined via an approximation process, and at the end we are
able to fix on the definitive value (denoted by $F_\infty$) which we shall use to obtain
a contradiction.
Second, our analysis as a whole generates enough `neighbourhood conditions' on $G_0$
to secure the entire computation: that is, we obtain a certain neighbourhood
$\Gcal \subseteq \SP^0(2)$ containing $G_0$ such that for any $G \in \Gcal$
we have $\Psi_0 \cdot F_\infty \cdot G = c$. 
The definition of $\Gcal$ together with this key property are established in Section~\ref{3.4}.

The remainder of the proof proceeds along the lines already indicated.
In Section~\ref{sec-ctex}, we draw on the above analysis to construct a certain procedure $G_1$ 
designed to force $\Phi_0$ to explore parts of $\TT(F_\infty)$ beyond the reach of $\Psi_0$.
In Section~\ref{sec-properties} we verify the required properties of $G_1$,
namely that $G_1 \in \Gcal$ (so that $\Psi_0 \cdot F_\infty \cdot G_1 = c$)
and also that \( \Phi_0 \cdot F_\infty \cdot G_1\) yields some value $K$ different from $c$.
Since this latter fact will hold for \emph{any} genuine bar recursor $\Phi$,
this establishes that $\Psi$ is not a genuine bar recursor.

We now proceed to the formal details of the proof.

\subsection{Computation analysis: the top level}  \label{3.1}

As indicated above, we begin by supposing that $\Phi,\Psi \in \SP^0(2 \arrow 2 \arrow 1)$
are simplified weak Kohlenbach bar recursors in the sense of Definition~\ref{restricted-def},
and assuming for contradiction that $\Psi$ is LWF.
We set $\Phi_0 = \lambda FG.\, \Phi F G \ang{} \in \SP^0(2 \arrow 2 \arrow 0)$
(or more formally $\Phi_0 = \lambda FG.\, \Phi \cdot F^\eta \cdot G^\eta \cdot \ang{}$),
and similarly $\Psi_0 = \lambda FG.\, \Psi F G \ang{}$.
Clearly $\Psi_0$ is LWF by  Theorem~\ref{substructures-thm}(i).

In general, if $t$ is any NSP term possibly containing $F^2, G^2$ free, 
and $F',G' \in \SP^0(2)$, we shall write $t[F',G']$ for the closed term obtained
from $t$ by instantiating $F,G$ to $F',G'$ and then evaluating.
(For instance, if $t$ is a procedure then formally $t[F',G'] = (\lambda FG.t) \cdot F' \cdot G'$.)

To start our construction, we consider the $\sysT_0^\str$-definable procedures
\begin{eqnarray*}
G_0   & = & \lambda g.\,\caseof{g(0)}{i \darrow 2i}      \;, \\
F^+_0 & = & \lambda f.\,\caseof{f(0)}{i \darrow \ang{i}} \;.
\end{eqnarray*}
The purpose of the doubling in the definition of $G_0$ is hard to motivate here,
but will emerge in Section~\ref{sec-properties}.
The functional $F^+_0$ represents a very simple well-founded tree:
note that $\ang{}$ is not a leaf in $\TT(F^+_0)$, but $\ang{x_0}$ is a leaf for every $x_0 \in \N$.

The definition of simplified weak bar recursor now implies that
$\Psi_0 \cdot F^+_0 \cdot G_0$ now evaluates to a certain $c \in \N$, or more formally to $\lambda.c$.
(In fact $c = 4 \ang{0} + 2$, but we will not need this information.)
By continuity of application, we may pick $k^0 > 0$ large enough that
$\Psi_0 \cdot F_0 \cdot G_0 = c$, where
\[ F_0 ~=~ \lambda f.\,
   \caseof{f(0)}{i<k^0 \darrow \ang{i} \mid i \geq k^0 \darrow \bot} \]
(extending our notation for $\ccase$ expressions in an obvious way).
Note that $F_0 \sqsubseteq F^+_0$.
We shall actually use $F_0$ (rather than $F^+_0$) as the first step in our approximative construction
of a suitable $F$.

Let us now look at the computation of $\Psi_0 \cdot F_0 \cdot G_0 = c$.
This will take the form of a head reduction of $\Psi_0 F_0 G_0$, 
and by inspection of the reduction rules in Section~\ref{sec-NSPs},
it is clear that this will follow a path through the syntax tree of $\Psi_0$ 
consisting of a finite sequence of calls to $F$ or $G$ (in any order), and leading to a leaf $c$.
For example, such a path might have the form
\[ \lambda FG.\; \ccase~ F(f^0_0) ~\oof~ u^0_0 \darrow 
   \ccase~G(g^0_0) ~\oof~ v^0_0 \darrow
   \ccase~F(f^0_1) ~\oof~ u^0_1 \darrow \cdots \darrow c \;.
\]
where the $f^0_i$ and $g^0_i$ are themselves type 1 procedures which 
appear syntactically within $\Psi_0$ and which may contain $F,G$ as free variables
(the superscript indicates that we are here analysing the computation at `level 0').
We can view the tracing of such a path through $\Psi$ as the `top level computation';
in addition to this, there will be subcomputations showing (for instance) 
that $(F f^0_0)[F_0,G_0]$ evaluates to $u^0_0$
and $(G g^0_0)[F_0,G_0]$ evaluates to $v^0_0$,

Let $f^0_0,\ldots,f^0_{l^0-1}$ be the complete list of such procedures
appearing as arguments to $F$ along this computation path,
with $u^0_0,\ldots,u^0_{l^0-1}$ the corresponding outcomes when $F,G$ are instantiated to $F_0,G_0$.
Likewise, let $g^0_0,\ldots,g^0_{n^0-1}$ be the list of procedures
appearing as arguments to $G$ on this path, with $v^0_0,\ldots,v^0_{n^0-1}$
the corresponding outcomes.

Of course, when $F=F_0$ and $G=G_0$, the procedures $f^0_i$ and $g^0_i$ will be interrogated 
only on the argument 0. This suggests that in order to `secure' the whole computation,  
we will also want to analyse the computations of 
$(f^0_i 0)[F_0,G_0]=u^0_i$ and $(g^0_ i 0)[F_0,G_0]=v^0_i$ for each $i$.
In fact, we shall do more: in order to give ourselves sufficient room for manoeuvre to construct 
the contrary example $G_1$ below (with the assurance of a similar evaluation behaviour for $G_1$),
we shall analyse the behaviour of each $g^0_i$ on all integer arguments $z$ up to a certain 
\emph{modulus} $m^0$.
In fact, it will suffice to take
\[ m^0 > k^0 + n^0 + 1 \;. \]
Again, this condition is hard to motivate at this stage;
the reason for it will emerge during the construction of $G_1$ in Section~\ref{sec-ctex},
at the point where we select $x_0$, the first step in our critical path through $\TT(F_\infty)$.

In order to proceed further, we need to extend our approximation to $F$.
First, extend the procedure $F_0$ to a $\sysT_0^\str$-definable $F^+_1$:
\begin{eqnarray*}
F^+_1 & =~ \lambda f. &
    \ccase~ {f(0)} ~\oof~ (i_0<k^0 \darrow \ang{i_0} 
                           \mid i_0 \geq k^0 \darrow \\
& & \caseof{f(1)}{i_1 \darrow \ang{i_0,i_1}}) \;.
\end{eqnarray*}
(It is an easy exercise to verify that this is indeed $\sysT_0^\str$-definable.)
The idea is that $\ang{x_0}$ will be a leaf node in $\TT(F^+_1)$ 
when $x_0 < k^0$, but elsewhere $\TT(F^+_1)$ will have depth 2.

We now consider the computation of $\Psi_0 \cdot F^+_1 \cdot G_0$.
Since $F^+_1 \sqsupseteq F_0$, this follows the same path through $\Psi_0$ as before
and features syntactically the same type 1 procedures $f^0_i$ and $g^0_i$ 
and the same outcomes $u^0_i,v^0_i$. 
Furthermore:

\begin{lemma}  \label{definedness-lemma}
(i) For any $i < n^0$ and any $z \in \N$, the evaluation of $g_i^0 [F^+_1,G_0] \cdot z$ yields 
a natural number, which we denote by $r^0_{iz}$.

(ii) For any $i < l^0$ and any $z \in \N$, the evaluation of $f_i^0 [F^+_1,G_0] \cdot z$ yields
a natural number, which we denote by $q^0_{iz}$.
\end{lemma}

\noindent \proof 
(i) Suppose for contradiction that 
$g^0_i[F^+_1,G_0](z) = \bot$ for some $i,z$, 
and let $G'_0 = \lambda g.\;\caseof{g(z)}{j \darrow G_0(g)}$.
Clearly $G'_0$ is $\sysT_0^\str$-definable.
Also $G'_0 \preceq G_0$ in the extensional preorder on NSPs,
so by Theorem~\ref{NSP-context-lemma} we have
\[ (G_0(g^0_i)) [F^+_1,G'_0] ~\sqsubseteq~ G'_0(g^0_i[F^+_1,G_0]) ~=~ \bot \;. \]
Moreover, for each application $F(f^0_j)$ 
(respectively $G(g^0_j)$) occurring before $G(g^0_i)$ 
in the path in question, 
we have $(F(f^0_j))[F^+_1,G'_0] \sqsubseteq u^0_j$
(respectively $(G(g^0_j))[F^+_1,G'_0] \sqsubseteq v^0_j$),
whence it is clear that $\Psi_0 \cdot F^+_1 \cdot G'_0$ is undefined.
But this contradicts Proposition~\ref{simplified-weak-br-defined-prop},
since both $F^+_1, G'_0$ are $\sysT_0^\str$-definable, $\TT(F^+_1)$ is well-founded 
and $\ang{\,} \in \TT(F^+_1)$.

The proof of (ii) is precisely similar.
\QED

\vspace*{1.5ex}
We shall make use of part~(i) of the above lemma for all $z < m^0$, and of part~(ii) only when $z=0$.
(The apparently superfluous use of $z$ in the latter case is intended to mesh with a more general
situation treated below.)

We may now make explicit some significant properties of $G_0$ in the form of \emph{neighbourhoods},
using the information gleaned so far.
For each $i<n^0$, define
\begin{eqnarray*} 
V^0_i & = & 
\{ g \in \SP^0(1) ~\mid~ 
   \forall z<m^0.\, g \cdot z = r^0_{iz} \}  \;, \\
\Gcal^0_i & = &
\{ G \in \SP^0(2) ~\mid~ \forall g \in V^0_i.\, G \cdot g = v^0_i \} \;.
\end{eqnarray*}
Clearly $G_0 \in \Gcal^0_i$ for each $i$, because $G_0$ interrogates 
its argument only at $0$. 
Also $g^0_i[F_1^+,G_0] \in V^0_i$ for each $i<n^0$.
This completes our analysis of the computation at top level;
we shall refer to this as the \emph{depth $0$ analysis}.

The idea is that the sets $\Gcal^0_i$ will form 
part of a system of neighbourhoods 
recording all the necessary information about $G_0$; we will then be free 
to select any $G_1$ from the intersection of these neighbourhoods 
knowing that the computation will proceed as before.
As things stand, the neigbourhoods $\Gcal^0_i$ do not achieve this:
for an arbitrary $G$ in all these neigbourhoods, there is no guarantee that
the value of each $g^0_i(z)$ at $F^+_1$ and $G$ will agree with its value
at $F^+_1$ and $G_0$ (and similarly for each $f^0_i(0)$).
To secure the whole computation, 
we therefore need a deeper analysis of these subcomputations in order to
nail down the precise properties of $G_0$ on which these rely.

\subsection{Computation analysis: the step case}  \label{3.2}

The idea is that we repeat our analysis for each of the finitely many computations of
\[ f^0_i(z) [F^+_1,G_0] ~~(i<l^0,\; z<1) \;, ~~~~~~~~
   g^0_i(z) [F^+_1,G_0] ~~(i<n^0,\; z<m^0) \;. \]
The analysis at this stage is in fact illustrative of the general analysis
at depth $w$, assuming we have completed the analysis at depth $w-1$.
For notational simplicity, however, we shall concentrate here on the
depth $1$ analysis, adding a few brief remarks on the depth $2$ analysis
in order to clarify how the construction works in general.

First, since each of the above computations yields a numeral $q^0_{iz}$ 
or $r^0_{iz}$ as appropriate, we may choose $k^1$  
such that all these computations yield the same results when $F^+_1$ is 
replaced by
\begin{eqnarray*}
F_1 & =~ \lambda f. &
    \ccase~ {f(0)} ~\oof~ (i_0<k^0 \darrow \ang{i_0} 
                           \mid i_0 \geq k^0 \darrow \\
& & \caseof{f(1)}{i_1<k^1 \darrow\ang{i_0,i_1} \mid i_1 \geq k^1 \darrow\bot}) 
\;. 
\end{eqnarray*}
Note in passing that $F_0$ no longer suffices here: there will be
computations of values for $g^0_i(z)$ that did not feature anywhere
in the original computation of $\Psi_0 \cdot F_0 \cdot G_0$.

Everything we have said about the main computation and its subcomputations
clearly goes through with $F^+_1$ replaced by $F_1$.
So let us consider the shape of the computations of
\[ f^0_i(z) [F_1,G_0] ~~(i<l^0,\; z<1) \;, ~~~~~~~~
   g^0_i(z) [F_1,G_0] ~~(i<n^0,\; z<m^0) \;. \]
At top level, each of these consists of a finite sequence of applications
of $F_1$ and $G_0$ (in any order), leading to the result 
$q^0_{iz}$ or $r^0_{iz}$.
Taking all these computations together, let 
$f^1_0,\ldots,f^1_{l^1-1}$ and $g^1_0,\ldots,g^1_{n^1-1}$ 
respectively denote the (occurrences of) type 1 procedures
to which $F_1$ and $G_0$ are applied, 
with $u^1_0,\ldots,u^1_{l^1-1}$ and $v^1_0,\ldots,v^1_{n^1-1}$
the corresponding outcomes.
Although we will not explicitly track the fact in our notation,
we should consider each of the $f^1_j$ and $g^1_j$ 
as a `child' of the procedure $f^0_i$ or $g^0_i$ from which it arose. 
Note that if $g^1_j$ is a child of $f^0_i$ 
(for example), then just as $F f^0_i$ appears as a subterm within the 
syntax tree of $\Psi_0$, so $G g^1_j$ appears as a subterm within the syntax
tree of $f^0_i$. Thus, each of the $f^1_j$ and $g^1_j$ corresponds to a
path in $\Psi_0$ with at least two left branches.

We now select a suitable modulus for our analysis of the $g^1_i$. Choose 
\[ m^1 > k^1 + n_0 + n_1 + 2 \;, ~~~~ m^1 \geq m^0 \;. \] 
(Again, the reason for this choice will emerge in Section~\ref{sec-ctex}.)
Extend $F_1$ to the $\sysT_0^\str$-definable functional 
\begin{eqnarray*}
F^+_2 & =~ \lambda f. &
    \ccase~ {f(0)} ~\oof~ (i_0<k^0 \darrow \ang{i_0} 
                           \mid i_0 \geq k^0 \darrow \\
& & \ccase~ {f(1)}~ \oof~ (i_1<k^1 \darrow \ang{i_0,i_1} 
                           \mid i_1 \geq k^1 \darrow \\
& & \ccase~ {f(2)}~ \oof~ (i_2 \darrow \ang{i_0,i_1,i_2}\,))) \;.
\end{eqnarray*}
Replacing $F_1$ by $F^+_2$ preserves all the structure established so far,
and just as in Lemma~\ref{definedness-lemma} 
we have that $g^1_i(z)$ at $F^+_2, G_0$
yields a numeral $r^1_{iz}$ for each $i<n^1$ and $z<m^1$;
similarly $f^1_i(z)$ at $F^+_2, G_0$ yields a numeral $q^1_{iz}$ for each
$i<l^1$ and $z<2$.
(In fact, it is superfluous to consider $f^1_i(1)$ in cases where
$f^1_i(0) < k^0$, 
but it simplifies notation to use $2$ here as our uniform modulus of inspection
for the $f^1_i$.)
We may now augment our collection of neighbourhoods by defining
\begin{eqnarray*} 
V^1_i & = & 
\{ g \in \SP^0(1) ~\mid~ 
   \forall z<m^1.\, g \cdot z = r^1_{iz} \} \;, \\
\Gcal^1_i & = &
\{ G \in \SP^0(2) ~\mid~ \forall g \in V^1_i.\, G \cdot g = v^1_i \} \;.
\end{eqnarray*}
for each $i<n^1$;
note once again that $G_0 \in \Gcal^1_i$ and that $g_1^0 [F_2^+,G_0] \in V^1_i$ for each $i$.
This completes our analysis of the computation at depth~1.

At the next stage,
we choose $k^2$ so that the above all holds with $F^+_2$ replaced by
\begin{eqnarray*}
F_2 & =~ \lambda f. &
    \ccase~ {f(0)} ~\oof~ (i_0<k^0 \darrow \ang{i_0} 
                           \mid i_0 \geq k^0 \darrow \\
& & \ccase~ {f(1)}~ \oof~ (i_1<k^1 \darrow \ang{i_0,i_1} 
                           \mid i_1 \geq k^1 \darrow \\
& & \ccase~ {f(2)}~ \oof~ (i_2<k^2 \darrow \ang{i_0,i_1,i_2} 
                           \mid i_2 \geq k^2 \darrow \bot))) \;.
\end{eqnarray*}
We now repeat our analysis for each of the computations of
\[ f^1_i(z) [F_2,G_0] ~~(i<l^1,\; z<2)   \;, ~~~~~~~~
   g^1_i(z) [F_2,G_0] ~~(i<n^1,\; z<m^1) \;. \]
Having identified the relevant type 1 procedures
$f^2_0,\ldots,f^2_{l^2-1}$ and $g^2_0,\ldots,g^2_{n^2-1}$ 
that feature as arguments to $F$ and $G$, we pick
\[ m^2 > k^2 + n^0 + n^1 + n^2 + 3 \;, ~~~~ m^2 \geq m^1 \;, \]
and use this to define suitable sets $V^2_i, \Gcal^2_i$ for $i < n^2$.
By this point, it should be clear how our construction may be continued to arbitrary depth.

\subsection{Computation analysis: the bottom level}  \label{3.3}

The crucial observation is that this entire construction eventually 
bottoms out.
Indeed, using $h$ as a symbol that can ambivalently mean either $f$ or $g$
(and likewise $H$ for $F$ or $G$),
we have that for any sequence $h^0_{i^0}, h^1_{i^1}, \ldots$ 
of type 1 procedures where each $h^{w+1}_{i^{w+1}}$ is a child of $h^w_{i^w}$,
the syntax tree of $\Psi_0$ contains the descending sequence of subterms
$H^0 h^0_{i^0}, H^1 h^1_{i^1}, \ldots$.
Since $\Psi_0$ is LWF by assumption, any such sequence must eventually terminate.
Moreover, the tree of all such procedures $h^w_i$ is finitely branching,
so by K\"onig's lemma it is finite altogether.

Let us see explicitly what happens at the last stage of the construction.
For some depth $d$, we will have constructed the $f^d_i, g^d_i, u^d_i, v^d_i$
as usual, along with $m^d$, $F^+_{d+1}$, the numbers $r^d_{iz}, q^d_{iz}$
and the neighbourhoods $\Gcal^d_i$,
but will then discover that $l^{d+1} = n^{d+1} = 0$: that is,
none of the relevant computations of $f^d_i(z)$ or $g^d_i(z)$
(relative to $F^+_{d+1}$ and $G_0$) themselves perform calls to $F$ or $G$.

At this point, we may settle on $F^+_{d+1}$ as the definitive version of $F$
to be used in our counterexample, and henceforth call it $F_\infty$.
Explicitly:
\begin{eqnarray*}
F_\infty & =~ \lambda f. &
    \ccase~ {f(0)} ~\oof~ (i_0<k^0 \darrow \ang{i_0} 
                           \mid i_0 \geq k^0 \darrow \\
& & \ccase~ {f(1)}~ \oof~ (i_1<k^1 \darrow \ang{i_0,i_1} 
                           \mid i_1 \geq k^1 \darrow \\
& & \cdots \\
& & \ccase~ {f(d)}~ \oof~ (i_d<k^d \darrow \ang{i_0,\cdots,i_d} 
                           \mid i_d \geq k^d \darrow \\
& & \ccase~ {f(d+1)}~ \oof~ (i_{d+1} \darrow \ang{i_0,\cdots,i_{d+1}}\,))
                           \cdots ))
\;.
\end{eqnarray*}
Clearly $F_\infty$ is $\sysT_0^\str$ and $F_\infty \sqsupseteq F_w$ for
$w \leq d$.
It is also clear from the above definition how $F_\infty$ represents a certain well-founded tree
$\TT(F_\infty)$ of depth $d+2$.
Note that if $f(0) \geq k^0, \ldots, f(d) \geq k^d$ then
$F_\infty \cdot f = \ang{f(0),\ldots,f(d+1)}$;
indeed $\ang{f(0),\ldots,f(d+1)}$ is a leaf in $\TT(F)$.
It is this portion of the tree, not visited by any of the computations
described so far, that we shall exploit when we construct our
counterexample $G_1$.

\subsection{The critical neighbourhood of $G_0$}  \label{3.4}

We may now define the \emph{critical neighbourhood} 
$\Gcal \subseteq \SP^0(2)$ by
\[  \Gcal ~=~ \bigcap_{w \leq d,~ i < n^w} \Gcal^w_i \;.  \]
Clearly $G_0 \in \Gcal$ by construction. 
Moreover, the following lemma shows that $\Gcal$ provides enough constraints to secure
the result of the entire computation:

\begin{lemma}  \label{secures-lemma}
For all $G \in \Gcal$ and all $w \leq d$, we have:
\begin{enumerate}
\item $f^w_i(z) [F_\infty,G] = q^w_{iz}$ 
      for all $i<l^w$ and $z \leq w$.
\item $g^w_i(z) [F_\infty,G] = r^w_{iz}$ 
      for all $i<n^w$ and $z < m^w$.
\item $F(f^w_i) [F_\infty,G] = u^w_i$
      for all $i<l^w$.
\item $G(g^w_i) [F_\infty,G] = v^w_i$
      for all $i<n^w$.
\item $\Psi_0 \cdot F_\infty \cdot G = c$.
\end{enumerate}
\end{lemma}

\noindent \proof
We prove claims 1--4 simultaneously by downwards induction on $w$.
For $w=d$, claims 1 and 2 hold because the computations in question make no
use of $F_\infty$ or $G$.
For any $w$, claim 1 implies claim 3: $F_w$ was chosen so that 
(among other things) $F_w (f^w_i[F_w,G_0]) = u^w_i$ is defined; 
moreover, $F_w$ interrogates its argument only on $0,\ldots,w$ at most,
so the established values of $f^w_i(z)[F_\infty,G]$ for $z \leq w$
suffice to ensure that $F_w(f^w_i[F_\infty,G]) = u^w_i$, and hence that
$F_\infty(f^w_i[F_\infty,G]) = u^w_i$.
Likewise, claim 2 implies claim 4, since
$G \in \Gcal^w_i$ by hypothesis, and the established values of
$g^w_i$ secure that $g^w_i \in V^w_i$ (at $F_\infty$ and $G$).

Assuming claims 3 and 4 hold for $w+1$, it is easy to see that
claims 1 and 2 hold for $w$: the relevant top-level computation may
be reconstituted from left to right leading to the result $q^w_{iz}$
or $r^w_{iz}$. Applying the same argument one last time also yields claim 5.
\QED

\subsection{The counterexample $G_1$}  \label{sec-ctex}

It remains to construct our contrary example $G_1 \in \Gcal$.
The idea is that $G_1$ will be chosen so that according to the definition
of simplified bar recursion, some value $K \neq c$ 
will be generated at depth $d+1$ of the tree $\TT(F_\infty)$ 
and then propagated up to the surface of the computation via nested applications of $G_1$.
We work with paths beyond the horizon defined by $k^0,k^1,\ldots$
to ensure that we do not encounter a leaf prematurely,
and also exploit the choice of moduli $m^w$ to ensure that the type 1 functions at intermediate
levels steer clear of the sets $V^w_i$.

Recall that $\Phi$ is assumed to be a genuine simplified bar recursor within $\SP^0$.
Set $\phi_0 = \Phi \cdot F_\infty \cdot G_0 \in \SP^0(1)$.
If $x$ is any sequence code $\ang{x_0,\ldots,x_{n-1}}$ and $z \in \N$, we shall write
$x.z$ for the sequence code $\ang{x_0,\ldots,x_{n-1},z}$, so that $x.z$ is the number 
computed by $\append \cdot x \cdot z$.

Since $G_0 = \lambda g.2g(0)$
and the leaf function has been fixed at $x \mapsto 2x+1$,
we have that for any sequence code $x$,
$\phi_0 \cdot x$ will take one of the values
\[ 2x+1 \;,~~ 2(2(x.0)+1) \;,~~ 4(2(x.0.0)+1) \;,~~ 
   8(2(x.0.0.0)+1) \;,~~ \ldots \;, \] 
according to where a leaf of $\TT(F_\infty)$ appears in the sequence $x,\, x.0,\, x.0.0,\, \ldots$.
In particular, for any fixed $j$, if we know that $j < |x|$,
we can recover $x_j$ from $\phi_0 \cdot x$ and even from $\theta(\phi_0 \cdot x)$,
where $\theta(n)$ denotes the unique odd number such that $n = 2^t.\theta(n)$
for some $t$. We shall write $x.0^t$ for the result of appending $t$
occurrences of $0$ to the sequence number $x$;
note that $\theta(\phi_0 \cdot x)$ will have the value $2(x.0^t)+1$ for some $t$.  

We construct a finite path $x_0,x_1,\ldots,x_d$ through the tree for $F_\infty$
in the following way, along with associated numbers $y_0,y_1,\ldots,y_d,y_{d+1}$.
Start by setting $y_0 = \phi_0 \cdot \ang{0}$.
Next, note that the mappings $z \mapsto \phi_0 \cdot \ang{z,0}$
and $z \mapsto \theta(\phi_0 \cdot \ang{z,0})$ are injective;
so because $m^0 > k^0 + n^0 + 1$, we may pick $x_0$ with $k^0 \leq x_0 < m^0$ such that:
\begin{itemize}
\item $y_1 = \phi_0 \cdot \ang{x_0,0}$ differs from $g^0_i(0)$ (more precisely from $r^0_{i0}$) 
for each $i<n^0$,
\item $\theta(y_1) = \theta(\phi_0 \cdot \ang{x_0,0})$ differs from $\theta(y_0)$.
\end{itemize}
Likewise, the mapping $z \mapsto \theta(\phi_0 \cdot \ang{x_0,z,0})$ is injective,
so since $m^1 > k^1 + n^0 + n^1 + 2$ we may pick $x_1$ with $k^1 \leq x_1 < m^1$ such that
\begin{itemize}
\item $y_2 = \phi_0 \cdot \ang{x_0,x_1,0}$ is different from all $r^0_{i0}$ and $r^1_{i'0}$ 
where $i<n^0$, $i'<n^1$, 
\item $\theta(y_2)$ is different from $\theta(y_0)$ and $\theta(y_1)$.
\end{itemize}
In general, we pick $x_w$ with $k^w \leq x_w < m^w$ such that 
\begin{itemize}
\item $y_{w+1} = \phi_0 \cdot \ang{x_0,\ldots,x_w,0}$ is different from all $r^u_{i0}$ 
with $u \leq w$ and $i<n^u$, 
\item $\theta(y_{w+1})$ is different from $\theta(y_0),\ldots,\theta(y_w)$. 
\end{itemize}
In each case, the first condition ensures that the type~1 function
$\Lambda z.\,\phi_0 \cdot \ang{\vec{x},z}$ steers clear of the sets $V^u_i$,
so that the functional $G_1$ to be defined below remains within $\Gcal$.
The second condition will ensure that the nested calls to $G_1$ do not interfere with one another
in their role of propagating the special value $K$.
Since $x_w \geq k^w$ for each $w \leq d$,
we have that $\ang{x_0,\ldots,x_d,0}$ is a leaf of $\TT(F_\infty)$.

We now take $K$ to be some natural number larger than any that has featured
in the construction so far, and in particular different from $c$, and define
\[ \begin{array}{rl}
G_1 ~=~ \lambda g. & \ccase~ g(0)~ \oof~ ( \\
     & y_{d+1} \darrow K \\
\mid & y_d \darrow \caseof{g(x_d)}{K \darrow K \mid j \darrow 2i} \\
\mid & \cdots \\
\mid & y_1 \darrow \caseof{g(x_1)}{K \darrow K \mid j \darrow 2i} \\
\mid & y_0 \darrow \caseof{g(x_0)}{K \darrow K \mid j \darrow 2i} \\
\mid & i \darrow 2i \\
) \;.
\end{array} \]
Here we understand $i,j$ as `pattern variables' that catch all cases
not handled by the preceding clauses.
In particular, the clauses $j \darrow 2i$, $i \darrow 2i$ mean that unless 
$g$ possesses some special property explicitly handled by some
other clause, we will have $G_1 \cdot g = 2(g \cdot 0) = G_0 \cdot g$.
It is straightforward to verify that $G_1$ is $\sysT_0^\str$-definable,
bearing in mind the availability of $\ifzero$ in $\sysT_0^\str$ 
(see the discussion in Section~\ref{sec-weaker-langs}).

\subsection{Properties of $G_1$}  \label{sec-properties}

We first check that $G_1$ falls within the critical neighbourhood:

\begin{lemma}  \label{G1-in-GG-lemma}
$G_1 \in \Gcal$, whence $\Psi \cdot F_\infty \cdot G_1 \cdot \ang{\,} = c$.
\end{lemma}

\noindent \proof
Suppose $w \leq d$ and $i < n^w$; we will show $G_1 \in \Gcal^w_i$.
Consider an arbitrary $g \in V^w_i$ (note that $g$ is not
assumed to represent a total function);
we want to show that $G_1 \cdot g = v_i^w$.
From the definition of $V^w_i$ we have $g \cdot 0 = r^w_{i0}$,
so for $u > w$, we have $g\cdot 0 \neq y_{u}$ by choice of $y_u$.
If also $g\cdot 0 \neq y_{u}$ for each $u \leq w$ then 
$G_1(g) = 2(g\cdot 0) = G_0(g) = v^w_i$ as required.
If $g\cdot 0 = y_{u}$ for some $u \leq w$,
then $G_1(g) = \caseof{g(x_{u})}{K \darrow K \mid i \darrow 2(g\cdot 0)}$.
However, since $x_{u} < m^{u} \leq m^w$  
we have $g(x_{u}) = r^w_{ix_{u}}$, and $K$ was assumed to be
larger than this, so once again $G_1(g) = 2(g\cdot 0) = v^w_i$.

By claim~5 of Lemma~\ref{secures-lemma}, 
it follows that $\Psi \cdot F_\infty \cdot G_1 \cdot \ang{\,} = c$.
\QED

\vspace*{1.5ex}
We now work towards showing that, by contrast, $\Phi \cdot F_\infty \cdot G_1 \cdot \ang{} = K$.
Set $\phi_1 = \Phi \cdot F_\infty \cdot G_1$, and
for $w \leq d+1$, denote $\ang{x_0,\ldots,x_{w-1}}$ by $x^w$.

\begin{lemma}
$\phi_1 \cdot (x^w.0) = y_w$ for all $w \leq d+1$.
\end{lemma}

\noindent \proof
Recall that $y_w = \phi_0 \cdot (x^w.0)$, which has the form $2^t.s$
where $s = \theta(y_w) = \phi_0 \cdot (x^w.0.0^t)$ and $x^w.0.0^t$ is a leaf for $F_\infty$,
so that $s = 2(x^w.0.0^t) + 1$.
Note also that if $0 < t' \leq t$ then $\phi_0 \cdot (x^w.0.0^{t'}) = 2^{t-t'}.s$,
which is distinct from $y_w$ (this is the point of the doubling in the definition of $G_0$),
and also from all the other $y_u$ since $\theta(y_0),\ldots,\theta(y_{d+1})$ are all distinct.

We may now see by reverse induction on $t' \leq t$ that
$\phi_1 \cdot (x^w.0.0^{t'}) = 2^{t-t'}.s$.
When $t'=t$, this holds because
$x^w.0.0^t$ is a leaf for $F_\infty$
so $\phi_1 \cdot (x^w.0.0^t) = \phi_0 \cdot (x^w.0.0^t)$.
Assuming this holds for $t'+1$ with $0 \leq t'<t$, 
because $x^w.0.0^{t'}$ is not a leaf we have
\begin{eqnarray*}
\phi_1 \cdot (x^w.0.0^{t'}) 
& = & G_1 \cdot (\lambda z. \phi_1 \cdot (x^w.0.0^{t'}.z)) \\
& = & \caseof{\phi_1 \cdot (x^w.0.0^{t'}.0)}{\cdots \mid i \darrow 2i} \\
& = & \caseof{2^{t-(t'+1)}.s}{\cdots \mid i \darrow 2i} \\
& = & 2^{t-t'}.s \;,
\end{eqnarray*}
using the observation that $2^{t-(t'+1)}.s$ is distinct from all of the $y_u$.

In particular, $\phi_1 \cdot (x^w.0) = 2^t.s = y_w$,
so the lemma is established.
\QED

\begin{lemma}
$\phi_1 \cdot x^w = K$ for all $0 \leq w \leq d+1$. 
\end{lemma}

\noindent \proof
By reverse induction on $w$.
For the case $w=d+1$, we have by the previous lemma that
$\phi_1\cdot (x^{d+1}.0) = y_{d+1}$, 
and since $x^{d+1}$ is not a leaf for $F_\infty$, we have
\begin{eqnarray*}
\phi_1 \cdot x^{d+1} 
& = & G_1 (\lambda z.\, \phi_1 \cdot (x^{d+1}.z)) \\
& = & \caseof{\phi_1 \cdot (x^{d+1}.0)}{y_{d+1} \darrow K \mid \cdots} \\
& = & K \;.
\end{eqnarray*}
For $w<d+1$, again we have by the previous lemma that $\phi_1 \cdot (x^w.0) = y_w$, 
and the induction hypothesis gives us
$\phi_1 \cdot (x^w.x_w) = \phi_1 \cdot (x^{w+1}) = K$.
Since $x^w$ is not a leaf for $F_\infty$, we have
\begin{eqnarray*}
\phi_1 \cdot x^w & = & G_1 (\lambda z.\, \phi_1 \cdot (x^w.z)) \\
   & = & \ccase~ {\phi_1 \cdot (x^w.0)}~ \oof \\
   &   & ~~~~(\cdots \mid 
              y_w \darrow \caseof{\phi_1 \cdot (x^w.x_w)}{K\darrow K \mid \cdots} 
              \mid \cdots) \\
   & = & K \;. ~~~~~~~~\mbox{\QED}
\end{eqnarray*}

\vspace*{1.5ex}
In particular, when $w=0$ we have $\phi_1 \cdot \ang{\,} = \phi_1 \cdot x^0 = K$.
Combining this with Lemma~\ref{G1-in-GG-lemma}, we have 
\[ \Psi \cdot F_\infty \cdot G_1 \cdot \ang{} ~=~  c ~\neq~ K ~=~ \phi_1 \cdot \ang{\,} ~=~
   \Phi \cdot F_\infty \cdot G_1 \cdot \ang{} \;.
\]
Since this argument applies for any genuine weak simplified bar recursor $\Psi$,
we may conclude that $\Psi$ is not a restricted bar recursor after all.
This completes the proof of Theorem~\ref{main-theorem}.

\section{Other models}   \label{sec-other-models}

Finally, we show how our non-definability result now transfers readily
to settings other than $\SP^0$, both partial and total.
The combined message of these results will be that bar recursion is not computable in 
$\sysT+\mmin$ or $\sysW$ in any reasonable sense whatever, 
however one chooses to make such a statement precise.

\subsection{Partial models}

It is relatively easy to transfer Theorem~\ref{main-theorem} to other `partial' settings,
by which we here mean simply-typed $\lambda$-algebras $\AA$ with $\AA(\nat) \iso \N_\bot$.
As a first step, it is convenient to detach our main theorem from $\SP^0$ and present its content
in purely syntactic terms.

Let $F:2$ be a closed term of $\sysT_0^\str$.
Using the $\sysT_0^\str$ program `$\basic$' introduced in Section~\ref{sec-bar-recursors},
and reinstating the hat notation for programming language numerals,
we may say a sequence $\vec{x} \in \N^*$ satisfies the Kohlenbach bar condition w.r.t.\ $F$
if the closed $\sysT_0^\str$ terms 
\[  F(\basic(\widehat{\ang{\vec{x}}}, \num{0})) \;, ~~~~~~
     F(\basic(\widehat{\ang{\vec{x}}}, \num{1}))  \]
evaluate to the same numeral;
we may thus define $\TT^K(F)$ to be the tree of sequences $\vec{x}$ 
such that no proper prefix of $\vec{x}$ satisfies this bar condition.
It is clear that this purely syntactic definition of $\TT^K(F)$
agrees with Definition~\ref{NSP-SK-trees} for NSPs:
if $\sem{F}$ is the denotation of $F$ in $\SP^0$, then $\TT^K(F) = \TT^K(\sem{F})$
by the adequacy of $\sem{-}$.

This allows us to reformulate the content of Theorem~\ref{main-theorem} syntactically as follows.
Here we write $=$ to mean that the closed programs on either side evaluate to the same numeral.

\begin{theorem}  \label{no-syntactic-BR-thm}
There is no closed $\sysT+\mmin$ term $\BR : 2 \arrow 2 \arrow 1$ such that the following hold
for all closed $\sysT_0^\str$ terms $F,G : 2$ with $\TT^K(F)$ well-founded, and for all 
$\vec{x} \in \TT^K(F)$:
\[ \begin{array}{rcl}
\BR\; F\, G\, \widehat{\ang{\vec{x}}} & = & 
   \widehat{2 \ang{\vec{x}}+1} \mbox{~~~if $\vec{x} \in \TT^K(F)^l$} \;, \\
\BR\; F\, G\, \widehat{\ang{\vec{x}}} & = & 
       G (\lambda z^\nat.\; \BR\; F\, G\, (\append \, \widehat{\ang{\vec{x}}} \, z))
        \mbox{~~~if $\vec{x} \in \TT^K(F)^n$} \;.
\end{array} \]
\end{theorem}

\noindent \proof
If such a term $\BR$ existed, then by adequacy of $\sem{-}$, $\Phi = \sem{\BR} \in \SP^0$ would be a 
$\sysT+\mmin$ definable simplified weak Kohlenbach recursor, 
contradicting Theorem~\ref{main-theorem}.
\QED

\vspace*{1.5ex}
Now suppose $\AA$ is any simply typed $\lambda$-algebra equipped with elements 
$0,1,\ldots$, $\suc$, $\pre$, $\ifzero$, $\rec_\sigma$, $\mmin$ of the appropriate types,
such that the induced interpretation $\sem{-}_\AA$ of $\sysT+\mmin$ in $\AA$ is \emph{adequate}:
that is, for closed programs $M:\nat$ and $n \in \N$, we have $\sem{M}_\AA = n \in \AA(\nat)$
iff $M \reducesto^* n$.
Note that this requires $\AA(\nat)$ to contain elements other than the numerals,
since in the presence of $\mmin$, diverging programs are possible.
In most cases of interest, we will have $\AA(\nat) \iso \N_\bot$:
typical examples include the Scott model $\PC$ of partial continuous functionals,
its effective submodel $\PC^\eff$, the model $\SF$ of PCF-sequential functionals
(arising as the extensional quotient of $\SP^0$) and its effective submodel $\SF^\eff$ of
PCF-computable functionals.

In this setting, we have a notion of $\sysT_0^\str$-definable element of $\AA$, 
so Definitions~\ref{NSP-SK-trees}, \ref{weak-bar-rec-def} and \ref{restricted-def} 
immediately relativize to $\AA$, giving us the notion of a (simplified) weak (Spector or Kohlenbach) 
bar recursor within $\AA$. We are now able to conclude:

\begin{theorem}
No simplified weak Kohlenbach bar recursor within $\AA$ can be $\sysT+\mmin$ definable.
\end{theorem}

\noindent \proof
If $\BR$ were a term of $\sysT+\mmin$ defining a simplified weak Kohlenbach bar recursor in $\AA$,
then by adequacy of $\sem{-}_\AA$, $\BR$ would satisfy the conditions in Theorem~\ref{no-syntactic-BR-thm},
a contradiction.
\QED

\vspace*{1.5ex}
The corresponding results for Spector bar recursion follow by Proposition~\ref{spect-kohlen-prop} 
relativized to $\AA$. 
It is also clear that we obtain similar results with $\sysW$ in place of $\sysT+\mmin$.

\subsection{Total models}

We now consider the situation for total type structures such as $\Ct$ and $\HEO$.
We work in the general setting of a simply-typed total combinatory algebra $\AA$ with $\AA(\nat)=\N$.

Our formulation for total models will have a character rather different from the above:
since no suitable element $\mmin$ will be present in $\AA$, we cannot induce an interpretation $\sem{-}$
straightforwardly from an interpretation of the constants---indeed, there will be terms of $\sysT+\mmin$
that have no denotation in $\AA$. Instead, we resort to an approach more in the spirit of Kleene's 
original definition of computability in total settings, adapting the treatment in \cite{HOC}.
It is best here to assume that $\AA$ is \emph{extensional}:
in fact, we shall assume that each $\AA(\sigma\arrow\tau)$ is a set of functions
$\AA(\sigma) \arrow \AA(\tau)$.
It is well-known that this implies that $\AA$ is a typed $\lambda$-algebra (see \cite[Section~4.1]{HOC}).
We shall furthermore assume that $\AA$ is a model of $\sysT_0$: that is, $\AA$ contains elements
$\suc$, $\pre$, $\ifzero$, and $\rec_\sigma$ for $\sigma$ of level 0
satisfying the usual defining equations for these constants.
(Note that in the total extensional setting, there is no real difference between $\sysT_0$ and $\sysT_0^\str$,
or between $\rec_\sigma$ and $\rec_\sigma^\str$, and the operator $\byval$ is redundant.
We shall henceforth use $\sysT_0$ in this context as it is directly a sublanguage of $\sysT+\mmin$.)

First, we recall that every $\sysT+\mmin$ term is $\beta\eta$-equivalent to one in 
\emph{long $\beta\eta$-normal form}---that is, to a $\beta$-normal term in which every
occurrence of any variable or constant $f$ is fully applied 
(i.e.\ appears at the head of a subterm $f N_0 \ldots N_{r-1}$ of type $\nat$). 
We shall define a (partial) interpretation in $\AA$ for $\sysT+\mmin$ terms of this kind,
and will in general write $\nf(M)$ for the long $\beta\eta$-normal form of $M$.%
\footnote{The correspondence between $\beta$-normal forms and Kleene-style indices is
explained in \cite[Section~5.1]{HOC}. Here we use long $\beta\eta$-normal forms in this role
because of our treatment of $\suc$ and $\rec_\sigma$ as first-class constants.}

For a given term $M$, a \emph{valuation} $\nu$ for $M$ will be a map assigning to each free variable
$x^\sigma$ within $M$ an element $\nu(x) \in \AA(\sigma)$.
We shall define a partial interpretation assigning to certain terms $M:\sigma$ and valuations $\nu$ for $M$
an element $\sem{M}_\nu \in \AA(\sigma)$.
This takes the form of an inductive definition of the relation $\sem{M}_\nu = a$, where $M$ is a 
$\beta\eta$-normal form of some type $\sigma$, $\nu$ is a valuation for $M$, and $a \in \AA(\sigma)$.
\begin{enumerate}
\item $\sem{\num{n}}_\nu = n$.
\item If $\sem{M}_\nu = n$ then $\sem{\suc\,M}_\nu = n+1$ and $\sem{\pre\,M}_\nu = n \dot{-} 1$,
where $\dot{-}$ is truncated subtraction.
\item If $\sem{M}_\nu = 0$ and $\sem{N}_\nu = n$, then $\sem{\ifzero\,M\,N\,P}_\nu = n$.
\item If $\sem{M}_\nu = m+1$ and $\sem{P}_\nu = n$, then $\sem{\ifzero\,M\,N\,P}_\nu = n$.
\item If $\sem{N}_\nu = 0$ and $\sem{\nf(X\vec{Y})}_\nu = m$, 
then $\sem{\rec_\sigma\,X\,F\,N\,\vec{Y}}_\nu = m$.
\item If $\sem{N}_\nu = n+1$ 
and $\sem{\nf(F\,(\rec_\sigma\,X\,F\,\num{n})\,\num{n}\,\vec{Y})}_\nu = m$, 
then $\sem{\rec_\sigma\,X\,F\,N\,\vec{Y}}_\nu = m$.
\item If $\sem{N}_\nu = n$ and $\sem{\nf(F\,\num{n})}_\nu = 0$, then $\sem{\mmin\,F\,N}_\nu = n$.
\item If $\sem{N}_\nu = n$, $\sem{\nf(F\,\num{n})}_\nu = i+1$ 
and $\sem{\mmin\,F\,\num{n+1}}_\nu = m$, then $\sem{\mmin\,F\,N}_\nu = m$.
\item If $f \in \AA(\sigma\arrow\tau)$ and $\sem{M}_{\nu [y \mapsto a]} = f(a)$ for all $a \in \AA(\sigma)$,
where $y^\sigma \not\in \dom\;\nu$, then $\sem{\lambda y.M}_\nu = f$.
\item If $\nu(x) = f$ and $\sem{P_i}_\nu = a_i$ for each $i<r$, then
$\sem{x P_0 \ldots P_{r-1}}_\nu = f (a_0,\cdots,a_{r-1})$.
\end{enumerate}
Other ways of treating the recursors $\rec_\sigma$ would be possible:
the definition chosen above errs on the side of generosity, in that it is possible e.g.\ for
$\sem{\rec_\sigma\,X\,F\,\num{0}\,\vec{Y}}_\nu$ to be defined even when $\sem{X}_\nu$ is not.
Note too that we are not assuming that all the System~$\sysT$ operators $\rec_\sigma$ are actually present 
in $\AA$---if they are not, there will of course be many System~$\sysT$ terms 
whose denotations in $\AA$ are undefined.

\begin{definition}  \label{Tmin-computable-def}
We say a partial function $\Phi : \AA(\sigma_0) \times\cdots\times \AA(\sigma_{r-1}) \parrow \N$ 
is \emph{$\sysT+\mmin$ computable} if there is a closed $\sysT+\mmin$ term 
$M : \nat$ with free variables among $x_0^{\sigma_0},\ldots,x_{r-1}^{\sigma_{r-1}}$
such that for all $a_0 \in \AA(\sigma_0),\ldots,a_{r-1} \in \AA(\sigma_{r-1})$ and all $n \in \N$ we have
\[  \sem{M}_{x_0 \mapsto a_0,\, \ldots,\, x_{r-1} \mapsto a_{r-1}} ~\kleq~ \Phi (a_0,\ldots,a_{r-1}) \;, \]
where $\kleq$ means Kleene equality.
\end{definition}

Comparing this with the treatment in \cite[Section~5.1]{HOC}, it is clear that if we restrict our language to
$\sysT_0 + \mmin$, the computable partial functions over $\AA$ obtained as above coincide exactly with
Kleene's $\mu$-computable partial functions.
We also note in passing that if the language is extended with the operator $\mathit{Eval}$ described in
\cite[Section~5.1]{HOC}, the computable partial functions are exactly the Kleene S1--S9 computable ones.

Next, we may adapt earlier definitions to say what it means to be a bar recursor with respect to $\AA$.
Note that since $\AA$ is a model of $\sysT_0^\str$, all functions $[\vec{x}\,j^\omega]$
as defined in Section~\ref{sec-bar-recursors} are present in $\AA(1)$.

\begin{definition}  \label{AA-BR-def}
(i) For any $F \in \AA(2)$, the \emph{Kohlenbach tree} $\TT^K(F)$ consists of all 
$\vec{x}$ such that no proper prefix $\vec{x}'$ of $\vec{x}$ satisfies
$F([\vec{x}'\,0^\omega]) = F([\vec{x}'\,1^\omega])$.

(ii) A partial function 
$\Phi : \AA(2) \times \AA(2) \times \AA(0) \parrow \N$ is a \emph{simplified Kohlenbach bar recursor} 
if for all $F,G \in \AA(2)$ with $\TT^K(F)$ well-founded, and for all $\vec{x} \in \TT^K(F)$, we have
\[ \begin{array}{rcl}
\Phi (F,G,\ang{\vec{x}}) & = & 2 \ang{\vec{x}}+1 \mbox{~~~if $\vec{x} \in \TT^K(F)^l$} \;, \\
\Phi (F,G,\ang{\vec{x}}) & = & G (\Lambda z.\, \Phi (F,G,\ang{\vec{x},z}))
        \mbox{~~~if $\vec{x} \in \TT^K(F)^n$} \;.
\end{array} \]
\end{definition}
We shall take it to be part of the meaning of the latter condition that the relevant function
$\Lambda z. \Phi (F,G,\ang{\vec{x},z})$ is indeed present in $\AA(1)$;
this is in effect a further hypothesis on $\AA$ which holds in all cases of interest.

We mention a few examples, all of which fall within the scope of Theorem~\ref{total-models-thm} below:
\begin{enumerate}
\item In the Kleene-Kreisel model $\Ct$,
it is the case for any $F \in \Ct(2)$ that $\TT^K(F)$ is well-founded,
and indeed a simplified bar recursor $\Phi$ is present within $\Ct$ itself, as an element of
$\Ct(2 \arrow 2 \arrow 0 \arrow 0)$.
It is known, furthermore, that such an element is Kleene S1--S9 computable (see \cite[Section~8.3]{HOC}).

\item By contrast, in the model $\HEO$, there are functionals $F$ such that $\TT^K(F)$ is not well-founded
(such functionals arise from the \emph{Kleene tree} as explained in \cite[Section~9.1]{HOC}),
and consequently it is not possible to find a total bar recursor within $\HEO$ itself.
Nonetheless, partial simplified bar recursors $\Phi : \HEO(2) \times \HEO(2) \times \HEO(0) \parrow \N$
as defined above do exist and are Kleene computable, by the same algorithm as for $\Ct$.
The situation is in fact precisely similar for the full set-theoretic model $\Set$:
thus, partial bar recursors in the spirit of Definition~\ref{bar-rec-def} are Kleene computable over $\Set$.

\item Another model of a quite different character is Bezem's type structure of
\emph{strongly majorizable functionals} \cite{Bezem-majorizable}, 
which has been found to be valuable in proof theory.
Here, as in $\Ct$, a bar recursor lives as a total object within the model itself---despite the presence of
\emph{discontinuous} type 2 elements in the model.
\end{enumerate}

We shall show that no simplified Kohlenbach bar recursor for $\AA$ can be $\sysT+\mmin$ computable
in the sense above. 
This will follow easily from Theorem~\ref{no-syntactic-BR-thm} once we have established the `adequacy'
of our partial interpretation $\sem{-}$.
This we do by means of a standard logical relations argument.
For each $\sigma$, let us define a relation $R_\sigma(M,a)$ between closed $\sysT+\mmin$ terms
$M:\sigma$ and elements $a \in \AA(\sigma)$ as follows:
\begin{itemize}
\item $R_\nat(M,m)$ iff $M \reducesto^* \num{m}$.
\item $R_{\sigma\arrow\tau}(M,f)$ iff for all $N:\sigma$ and $a \in \AA(\sigma)$, 
$R_\sigma(N,a)$ implies $R_\tau(MN,f(a))$.
\end{itemize}
We often omit the type annotations and may refer to any of the $R_\sigma$ as $R$.

\begin{lemma}  \label{R-lemma}
If $\sem{M}_\nu = a$ and $R(N_i,\nu(x_i))$ for all $x_i$ free in $M$,
then $R(M[\vec{x} \mapsto \vec{N}],a)$.
\end{lemma}

\noindent \proof
By induction on the generation of $\sem{M}_\nu = a$ via clauses 1--10 above.
The cases for clauses 1--4 are trivial, and those for clauses 5--8 are very straightforward,
using the fact that any term $M$ is observationally equivalent to $\nf(M)$ by the context lemma.

For clause 9, suppose we have $\sem{\lambda y^\sigma.M}_\nu = f \in \AA(\sigma\arrow\tau)$ 
arising from $\sem{M}_{\nu[y \mapsto a]} = f(a)$ for all $a \in \AA(\sigma)$,
and suppose also that $R(N_i,\nu(x_i))$ for all $x_i$ free in $\lambda x.M$.
We wish to show that $R_{\sigma\arrow\tau}((\lambda y.M)[\vec{x} \mapsto \vec{N}],f)$:
that is, that for all $P:\sigma$ and $a \in \AA(\sigma)$, 
$R_\sigma(P,a)$ implies $R_\tau((\lambda y.M)[\vec{x} \mapsto \vec{N}](P), f(a))$.
So suppose $R_\sigma(P,a)$. By assumption, we have $\sem{M}_{\nu[x \mapsto a]} = f(a)$
and $R(N_i\,\nu(x_i))$ for all $i$, so $R_\tau(M[\vec{x} \mapsto \vec{N}, y \mapsto P],\,f(a))$
by the induction hypothesis. The desired conclusion follows, since
\[ (\lambda y.M)[\vec{x} \mapsto \vec{N}](P) ~\reducesto~
    M[\vec{x} \mapsto \vec{N}, y \mapsto P]  \]
and it is easy to see by induction on types that if $Q \reducesto Q'$ and $R(Q',b)$ then $R(Q,b)$.

For clause 10, suppose we have $\sem{x_j \vec{P}}_\nu = f(\vec{a})$ arising from
$\nu(x_j)=f$ and $\sem{P_i}_\nu = a_i$ for each $i$, and suppose again that $R(N_i,\nu(x_i))$ for all $i$.
Writing $^*$ for the substitution $[\vec{x} \mapsto \vec{N}]$, we have
$(x_j \vec{P})^* = N_j \vec{P}^*$, so it will suffice to show that $R(N_j \vec{P}^*, f(\vec{a}))$.
But we have $R(N_j,\nu(x_j))$ where $\nu(x_j)=f$, 
and also $R(P_i^*,a_i)$ for each $i$ by the induction hypothesis,
so by definition of $R_\sigma$ where $\sigma$ is the type of $N_j$, 
we have $R(N_j \vec{P}^*, f(\vec{a}))$ as required.
\QED


\vspace*{1.5ex}
The converse to Lemma~\ref{R-lemma} is not true. For instance, 
if $M = f(\mmin\,(\lambda y.\num{1})\,\num{0})$, $N = \lambda x.\num{2}$ 
and $a = \Lambda x.2 \in \AA(1)$, then $R_1(N,a)$ and $M[f \mapsto N] \reducesto^* \num{2}$,
but $\sem{M}_{f \mapsto a}$ is undefined because $\mmin\,(\lambda y.\num{1})\,\num{0}$
receives no denotation.
In this sense, $\sysT+\mmin$ computability in a total model is a stricter condition than it would be 
in a partial model.  It is therefore not too surprising that no bar recursor for a total model can be
$\sysT+\mmin$ computable.

Recall that we are assuming that $\AA$ is a model of $\sysT_0$, in the sense that $\AA$ contains
suitable elements $\suc$, $\pre$, $\ifzero$, $\rec_\sigma$ satisfying the relevant equations,
giving rise via the $\lambda$-algebra structure of $\AA$ to an interpretation of $\sysT_0$ which we shall
denote by $I$. We may now verify that our two ways of interpreting $\sysT_0$ terms are in accord:

\begin{lemma}  \label{T_0-lemma}
Suppose $M$ is any long $\beta\eta$-normal $\sysT_0$ term, and $\nu = (\vec{x} \mapsto \vec{a})$
is any valuation for $M$. Then $\sem{M}_\nu$ is defined and is equal to $I_{\vec{x}}(M)(\vec{a})$.
\end{lemma}

\noindent \proof
A routine induction on the structure of $M$. 
\QED

\vspace*{1.5ex}
We now have all the pieces needed for the main result, which establishes Corollary~6.3.33 of \cite{HOC}.

\begin{theorem}  \label{total-models-thm}
No simplified Kohlenbach bar recursor for $\AA$ can be $\sysT+\mmin$ computable.
\end{theorem}

\noindent \proof 
Suppose $B$ were a $\sysT+\mmin$ term with free variables $F:2$, $G:2$, $x:0$ 
defining a simplified Kohlenbach bar recursor 
$\Phi : \AA(2) \times \AA(2) \times \AA(0) \parrow \N$ as above.
We claim that $\BR = \lambda FGx.B$ satisfies the conditions of Theorem~\ref{no-syntactic-BR-thm},
yielding a contradiction.
Indeed, suppose $\hat{F},\hat{G} : 2$ are closed $\sysT_0^\str$ terms 
with $\TT^K(\hat{F})$ well-founded.
Construing $\hat{F},\hat{G}$ as $\sysT_0$ terms, we obtain elements 
$\sem{\hat{F}}, \sem{\hat{G}} \in \AA(2)$ by Lemma~\ref{T_0-lemma},
and it is clear from Lemma~\ref{R-lemma} that $\TT^K(\sem{\hat{F}})$ = $\TT^K(\hat{F})$.

Now suppose $\vec{x} \in \TT^K(\hat{F})$.
We now show by meta-level bar induction on $\vec{x} \in \TT^K(\hat{F})$ that for all such $\vec{x}$,
$\sem{B}_{F \mapsto \sem{\hat{F}},\, G \mapsto \sem{\hat{G}},\, x \mapsto \ang{\vec{x}}}$
is defined and agrees with the value of $\BR\,\hat{F}\,\hat{G}\,\widehat{\ang{\vec{x}}}$, 
and moreover the latter satisfies the relevant condition of Theorem~\ref{no-syntactic-BR-thm}.

First, if $\vec{x} \in \TT^K(\hat{F})^l$, then by Definitions~\ref{Tmin-computable-def} and \ref{AA-BR-def}
we have
\[ \sem{B}_{F \mapsto \sem{\hat{F}},\, G \mapsto \sem{\hat{G}},\, x \mapsto \ang{\vec{x}}} ~=~ 
   \Phi\,(\sem{\hat{F}},\sem{\hat{G}},\ang{\vec{x}}) ~=~ 2 \ang{\vec{x}}+1 
\;. \]
Now by Lemma~\ref{R-lemma} we have $R(\hat{F}, \sem{\hat{F}})$,
$R(\hat{G}, \sem{\hat{G}})$ and $R(\widehat{\ang{\vec{x}}}, \ang{\vec{x}})$,
so by the same again we have
\( R_\nat\, (B[F \mapsto \hat{F}, G \mapsto \hat{G}, x \mapsto \num{\ang{\vec{x}}}],\, 
   2\ang{\vec{x}}+1) \),
meaning that 
\[ B[F \mapsto \hat{F}, G \mapsto \hat{G}, x \mapsto \num{\ang{\vec{x}}}]  ~\reducesto^*~
    \widehat{2\ang{\vec{x}}+1} \;. \]
Hence $\BR\,\hat{F}\,\hat{G}\,\widehat{\ang{\vec{x}}}$ 
satisfies the first condition of Theorem~\ref{no-syntactic-BR-thm}, and all parts of the induction claim
are established.

Now suppose that $\vec{x} \in \TT^K(\hat{F})^n$, where each child $\vec{x},z$ satisfies the induction claim.
We first show that
\[ R_{\nat\arrow\nat} ~ (\lambda z.\,\BR\;\hat{F}\;\hat{G}\;(\append\,\widehat{\ang{\vec{x}}}\;z), ~
             \Lambda z.\, \sem{B}_{F \mapsto \sem{\hat{F}},\, G \mapsto \sem{\hat{G}},\,
             x \mapsto \ang{\vec{x},z}}) \;. \]
For this, it suffices to show that if $z \in N$ and $R_\nat(Z,z)$ (i.e.\ $Z \reducesto^* \num{z}$), then
\[ R_\nat\; ((\lambda z.\,\BR\;\hat{F}\;\hat{g}\;(\append\,\widehat{\ang{\vec{x}}}\,z))Z,\; 
                 \sem{B}_{F \mapsto \sem{\hat{F}},\, G \mapsto \sem{\hat{G}},\,
             x \mapsto \ang{\vec{x},z}}) \;. \]
But this holds by the induction hypothesis along with the observational equivalence
\[  (\lambda z.\,\BR\;\hat{F}\;\hat{G}\;(\append\,\widehat{\ang{\vec{x}}}\,z))Z ~\obseq~
        \BR\;\hat{F}\;\hat{G}\;\widehat{\ang{\vec{x},z}} \;.  \]
        
Since $R(\hat{G},\sem{\hat{G}})$, we may conclude that
\[ R_\nat~ (\hat{G}\;(\lambda z.\,\BR\;\hat{F}\;\hat{G}\;(\append\,\widehat{\ang{\vec{x}}}\;z)),~ 
   \sem{\hat{G}}\;(\Lambda z.\, \sem{B}_{F \mapsto \sem{\hat{F}},\, G \mapsto \sem{\hat{G}},\,
             x \mapsto \ang{\vec{x},z}})) \]
so that the term on the left evaluates to (the numeral for) the value on the right.
But also by Definitions~\ref{Tmin-computable-def} and \ref{AA-BR-def} we have
\begin{eqnarray*}
\sem{B}_{F \mapsto \sem{\hat{F}},\, G \mapsto \sem{\hat{G}},\, x \mapsto \ang{\vec{x}}}
   & = & \Phi\,(\sem{\hat{F}}, \sem{\hat{G}}, \ang{\vec{x}}) \\
   & = & \sem{\hat{G}}\;(\Lambda z.\, \Phi\,(\sem{\hat{F}}, \sem{\hat{G}}, \ang{\vec{x},z})) \\
   & = & \sem{\hat{G}}\;(\Lambda z.\, \sem{B}_{F \mapsto \sem{\hat{F}},\, G \mapsto \sem{\hat{G}},\, x \mapsto \ang{\vec{x},z}})  \;.
\end{eqnarray*}
In particular, $\sem{B}_{F \mapsto \sem{\hat{F}},\, G \mapsto \sem{\hat{G}},\, x \mapsto \ang{\vec{x}}}$ 
is defined, so using Lemma~\ref{R-lemma} as before, we see that
$B[F \mapsto \hat{F}, G \mapsto \hat{G}, x \mapsto \num{\ang{\vec{x}}}]$ also evaluates to
$\sem{\hat{G}}\,(\Lambda z. \sem{B}_{F \mapsto \sem{\hat{F}},\, G \mapsto \sem{\hat{G}},\, x \mapsto \ang{\vec{x},z}})$.
Thus the induction claim is established for $\vec{x}$.

We have thus shown that $\BR$ satisfies the conditions of Theorem~\ref{no-syntactic-BR-thm},
so a contradiction with that theorem is established.
\QED

\vspace*{1.5ex}
Clearly, similar results hold for Spector bar recursion or for the language $\sysW$.

\end{document}